\documentclass[12pt]{article}
\begin{document}
\begin{large}
\title{\bf { A group of invariance transformations for nonrelativistic quantum mechanics }}
\end{large}
\author{Bruno Galvan \footnote{E-Mail: bgalvan@delta.it}\\ \small Loc. Melta 40, 38014 Trento, Italy.}
\date{\small February 2000}
\maketitle
\begin{abstract}
This paper defines, on the Galilean space-time, the group of {\it asymptotically Euclidean transformations} (AET), which are equivalent to Euclidean transformations at space-time infinity, and proposes a formulation of nonrelativistic quantum mechanics which is invariant under such transformations.
This formulation is based on the {\it asymptotic quantum measure}, which is shown to be invariant under AET's. This invariance exposes an important connection between AET's and Feynman path integrals, and reveals the {\it nonmetric} character of the asymptotic quantum measure. The latter feature becomes even clearer when the theory is formulated in terms of the coordinate-free formalism of {\it asymptotically Euclidean manifolds}, which do not have a metric structure.

This mathematical formalism suggests the following physical interpretation: (i) Particles evolution is represented by trajectories on an asymptotically Euclidean manifold; (ii) The metric and the law of motion are not defined {\it a priori} as fundamental entities, but they are properties of a particular class of reference frames; (iii) The universe is considered as a probability space in which the asymptotic quantum measure plays the role of a probability measure. Points (ii) and (iii) are used to build the {\it asymptotic measurement theory}, which is shown to be consistent with traditional quantum measurement theory. The most remarkable feature of this measurement theory is the possibility of having a {\it nonchaotic distribution of the initial conditions} (NCDIC), an extremely counterintuitive but not paradoxical phenomenon which allows to interpret typical quantum phenomena, such as particle diffraction and tunnel effect, while still providing a description of their motion in terms of classical trajectories. This paper also shows that in the presence of NCDIC, Bell's inequality can no longer be demonstrated.
\end{abstract}
\section{Introduction.}
The rise or full development of almost all major physical theories --such as classical mechanics, electromagnetism, special relativity theory and general relativity theory-- saw the introduction of a new group of transformations, with respect to which an invariant formulation of physical laws was required: Galilean transformations for classical mechanics, Lorentz-Poincar\'e transformations for electromagnetism and special relativity, and diffeomorphisms for general relativity.

Quantum mechanics is a remarkable exception to this rule, since no special group of transformations is associated with it. This observation is an incentive to search for such a group, in the hope that its discovery will allow to formulate the theory more elegantly and without the many paradoxes and problems that currently affect it.

With this aim in mind, the group of {\it asymptotically Euclidean transformations}, i.e., transformations which are equivalent to Euclidean transformations at space-time infinity, is defined on the Galilean space-time. A formulation of nonrelativistic quantum mechanics (Schr\"{o}dinger's equation) which is invariant with respect to such transformations is then proposed. The first part of this paper develops the corresponding mathematical theory; the second part discusses its physical interpretation. In the mathematical section, a demonstration will be provided only for the most significant lemmas and theorems.

This paper is a development a previous paper \cite{galvan}.


\section{Mathematical theory.}

\subsection{Asymptotically Euclidean transformations.}

For the sake of brevity, let $G$ designate the Galilean space-time $R\times R^3$. The transformations usually considered on $G$ are the Galilean transformations, which have the form $f(t,{\bf x})=(t+t_0,R{\bf x}-{\bf v}_0t+{\bf x}_0)$, where $R$ is an orthogonal matrix. The group of Galilean transformations is composed of three subgroups: translations $f(t, {\bf x})=(t+t_0,{\bf x}+{\bf x}_0)$, rotations $f(t, {\bf x})=(t,R{\bf x})$, and boosts $f(t, {\bf x})=(t, {\bf x}-{\bf v}_0t)$. Galilean transformations separately preserve space distances and time distances.

Given a set $A\subseteq G$, let $A(t)\subseteq R^3$ be the space cross-section of the set $A$ at the time $t$, i.e., $A(t):=\{ {\bf x}\in R^3|(t, {\bf x})\in A\}$. A {\it semitrajectory} $\gamma$  is the graph in $G$ of a continuous curve from $[t_0,+\infty )$ to $R^3$. In accordance with the above defined notation, $\gamma (t)\in R^3$ is the space point occupied by the semitrajectory at the time t. The point $(t_0,\gamma (t_0))$ is called origin of the semitrajectory. Clearly, this definition of semitrajectory implies that it is temporally headed into the future. This directionality, although undeclared, will always be implied in all the definitions, lemmas and theorems of this paper.

\subsubsection{ Asymptotic velocity.} \label{pippo}

Let us now introduce the important notion of asymptotic velocity of a semitrajectory. This notion, applied to classical trajectories, is already in use within classical scattering theory \cite{scattering}.

\newtheorem{definizione}{Definition}
\newtheorem{lemma}{Lemma}
\newtheorem{esempi}{Examples}
\newtheorem{teorema}{Theorem}
\newtheorem{corollario}{Corollary}
\newtheorem{congettura}{Conjecture}
\newtheorem{proposizione}{Proposition}

\begin{definizione} \rm 
(a) The {\it asymptotic velocity} of a semitrajectory is the limit of $\gamma (t)/t$ for $t\rightarrow +\infty$, when it exists.
(b) A semitrajectory allowing asymptotic velocity is said to be {\it asymptotically regular}.
\end{definizione}

Let $V$ be the space of asymptotic velocities (which is isomorphic with $R^3$); let $\Gamma$ be the set of asymptotically regular semitrajectories; and finally, let $\omega :\Gamma \rightarrow V$ be the application that associates each asymptotically regular semitrajectory with its asymptotic velocity.

\begin{lemma} \label{l:1} \rm
(a) If there is a ${\bf v}\in V$ such that $\|\gamma (t)- {\bf v}t\|$ is bounded for $t\geq t_0$, then $\omega (\gamma )={\bf v}$; in particular, if  $\|\gamma (t)\|$ is bounded, then $\omega (\gamma )=0$.
(b) If $\gamma (t)$ admits first derivative and $\dot{\gamma}(t)\rightarrow {\bf v}$ for $t\rightarrow +\infty$ , then $\omega (\gamma )= {\bf v}$.
(c) If $f(t, {\bf x})=(t+t_0,R{\bf x}-{\bf v}_0t+{\bf x}_0)$ is a Galilean transformation, then $\omega [f(\gamma )]=R\omega (\gamma )- {\bf v}_0$.
\end{lemma}

\begin{esempi} \label{e:1}\rm
(a) If $\gamma (t)= {\bf v}t+{\bf x}_0 \sin\omega t$, then $\omega (\gamma )= {\bf v}$.
(b) If $\gamma (t)= {\bf a}|t|^{(1-\epsilon )}+ {\bf x}_0$, with $0<\epsilon < 1$, then $\omega (\gamma )=0$.
(c) The curve $\gamma (t)= {\bf v}t \sin\omega t$ is not asymptotically regular.
\end{esempi}

The converse of Lemma \ref{l:1}a does not hold, i.e., it is possible to have $\omega (\gamma )=0$ while $\|\gamma (t)\|$ is not bounded: consider example \ref{e:1}c. Unbounded trajectories having vanishing asymptotic velocity are also termed {\it almost bounded trajectories}.

Hereafter, unless explicitly stated, the term trajectory is used to designate an asymptotically regular trajectory.


\subsubsection{Causal transformations.} \label{s212}

Let us now study homeomorphic transformations on $G$, for which we will use the more generic term {\it transformations}. This paper studies a group of transformations which is broader than Galilean transformations. One property that these transformation will still be required to have, however, is {\it causality}, i.e., they must maintain the time ordering of events. Let us specify this property by using first of all the following notation: if $(t,{\bf x})\in G$, we define the projectors  $P_T(t,{\bf x}):=t$ and $P_X(t,{\bf x}):={\bf x}$.

\begin{definizione} \rm
A transformation $f$ on $G$ is {\it causal} if, for every $u,w\in G$ such that  $P_T(u)<P_T(w)$, one has $P_T[f(u)]<P_T[f(w)]$.
\end{definizione}
\begin{lemma} \label{l:2} \rm
(a) If $f$ is causal, $P_T(u)=P_T(w)$ implies $P_T[f(u)]=P_T[f(w)]$.
(b) Causal transformations form a group.
(c) A causal transformation is of the form $f(t,{\bf x})=(f_T(t),f_X(t,{\bf x}))$, where $f_T:R\rightarrow R$ is a monotonically increasing homeomorphism, and $f_X(t,\cdot):R^3\rightarrow R^3$ is a homeomorphism for every $t$.
(d) If $f$ is causal and $A\subseteq G$, one has $f(A)[f_T(t)]=f_X[t,A(t)]$.
(e) A transformation $f$ is causal if and only if it preserves semitrajectories.
\end{lemma}

Hereafter, unless otherwise specified, the term {\it transformation} is used to designate a causal transformation.


\subsubsection{Asymptotic homeomorphisms.} \label{s213}

The following definitions meaningfully characterize the asymptotic behavior of a transformation.

\begin{definizione} \rm
(a) A transformation $f$ is {\it asymptotically regular at the point} ${\bf v}\in V$ if  there is a ${\bf v'}\in V$ such that for every semitrajectory $\gamma$  for which the relation $\omega (\gamma )={\bf v}$ holds, one has $\omega [f(\gamma )]= {\bf v'}$.
(b) A transformation $f$ is {\it asymptotically regular} if it is asymptotically regular in every point ${\bf v}\in V$.
(c) If $f$ is asymptotically regular, the application $f^+:V\rightarrow V$ defined by setting $f^+[\omega (\gamma )]:=\omega [f(\gamma )]$ is termed {\it asymptotic transform of f}.
(d)$f$ is an {\it asymptotic homeomorphism} if $f$ and $f^{-1}$ are asymptotically regular.
\end{definizione}

Clearly, the definition of asymptotic regularity can be considered the asymptotic equivalent of the usual definition of regularity of a function, which says that a function is regular at a point if it admits a limit at that point. The definition of asymptotic homeomorphism is also similar to the usual nonasymptotic definition.
\begin{esempi} \rm
(a) A Galilean transformation $f(t,{\bf x})=(t+t_0,R{\bf x}-{\bf v}_0t+{\bf x}_0)$ is an asymptotic homeomorphism, and one has $f^+({\bf v})=R{\bf v}-{\bf v}_0$. This is derived trivially from Lemma \ref{l:1}c. The asymptotic transform of a Galilean transformation is therefore a Euclidean transformation on the space of asymptotic velocities: rotations remain rotations, boosts become translations, and translations disappear.
(b) If $f(t,{\bf x})=(at,a{\bf x})$, where $a>0$ is a multiplicative constant, $f$ is asymptotically regular and the relation $f^+({\bf v})={\bf v}$ holds.
(c) The transformation $f(t,{\bf x})=(t, {\bf x}+{\bf v}_0t\sin\omega t)$ is not asymptotically regular at any point.
\end{esempi}

\begin{teorema} \label{t:1} \rm
(a) If $f$ is asymptotically regular, $f^+$ is continuous.
(b) If $f$ is an asymptotic homeomorphism, then $f^+$ is a homeomorphism on V and one has $(f^{-1})^+=(f^+)^{-1}$.
(c) If $f$ and $g$ are two asymptotic homeomorphisms, then $g\cdot f$ also is an asymptotic homeomorphism, and the relation $(g\cdot f)^+=g^+\cdot f^+$ holds.
\end{teorema}

In order to prove the above theorem it is convenient to modify the function $f$ so that asymptotic regularity is changed into normal regularity. In order to easily carry out this modification, instead of considering transformations on $G$, we will consider transformations on $G^+:=[a,+\infty )\times R^3\subseteq G$, where $a$ is a positive constant. Such a restriction is irrelevant to the remainder, since the theorem relate to the future asymptotic properties of  transformations.
Let us define $F^+:=(0,1/a]\times V$, $F_0:=\{0\}\times V$ and $F_0^+:=F^+\cup F_0$. If $(s,{\bf v})\in F_0^+$, let us define the projectors $P_S(s,{\bf v}):=s$ and $P_V(s,{\bf v}):={\bf v}$. Furthermore, let us define the application $h:G^+\rightarrow F^+$ by setting $h(t,{\bf x}):=(1/t, {\bf x}/t)$. The application $h$ is a homeomorphism between $G^+$ and $F^+$, and carries the time infinity of $G^+$ onto $F_0$. If $f$ is a transformation on $G^+$, we define $\hat{f}:F^+\rightarrow F^+$ by setting $\hat{f}:=h\cdot f\cdot h^{-1}$. The function $\hat{f}$ is a causal homeomorphism on $F^+$, i.e., $P_S(u)<P_S(w)$ implies $P_S[\hat{f}(u)]<P_S[\hat{f}(w)]$.

If $g$ is a homeomorphism on $F^+$ and is regular in $F_0$, we define $g_0:V\rightarrow V$ by setting $g_0({\bf v}):=P_V[\lim_{(s,{\bf v'})\rightarrow (0,{\bf v})}g(s,{\bf v'})]$; moreover, $P_S[\lim_{(s,{\bf v'})\rightarrow (0,{\bf v})}g(s,{\bf v'})]=0$, because the limit must be an accumulation point for $F^+$, but not belonging to it. The connection between the asymptotic behavior of a transformation $f$ on $G^+$ and the behavior of the transformation $\hat{f}$ on $F^+$ in the proximity of $F_0$ is described by the following lemma:
\begin{lemma} \label{l:3} \rm \rm $f$ is asymptotically regular if and only if $\hat{f}$ is regular in $F_0$, and the relation $f^+=\hat{f}_0$ holds.
\end{lemma}

Lemma \ref{03} allows us to obtain results regarding the asymptotic behavior of a homeomorphism on $G^+$ by studying the behavior of a homeomorphism on $F^+$ in the proximity of $F_0$. If $g$ is a homeomorphism on $F^+$ which is regular in $F_0$, we define $\bar{g}:F_0^+\rightarrow F_0^+$, setting
$$ \label{01} 
\bar{g}(s,{\bf v}):=\cases{g(s,{\bf v})$ if $s>0 \cr (0,g_0({\bf v}))$ if $s=0}.
$$
We have the following lemma:
\begin{lemma} \label{l:4} \rm
(a) If $g$ is a homeomorfism on $F^+$ and it is regular in $F_0$, then the functions $g_0$ and $\bar{g}$ are continuous.
(b) if $g$ and $g^{-1}$ are regular in $F_0$, then $g_0$ is invertible and the relation $(g_0)^{-1}=(g^{-1})_0$ holds; furthermore, $\bar{g}$ is a homeomorphism.
\end{lemma}
{\bf Proof}. We only give the proof that $g_0$ is continuous. Consider a sequence ${\bf v}_n\rightarrow {\bf v}_0\in V$. Due to the regularity of $g$ at every point $(0,{\bf v}_n)$, for every $n$ it is possible to find a point $(s_n,{\bf w}_n)\in F^+$ such that $\|(s_n,{\bf w}_n)-(0,{\bf v}_n)\|\leq 1/n$ and $\|g(s_n,{\bf w}_n)-(0,g_0({\bf v}_n))\|\leq 1/n$. Therefore $(s_n,{\bf w}_n)\rightarrow (0,{\bf v}_0)$ and $g(s_n,{\bf w}_n)\rightarrow (0,g_0({\bf v}_0))$. Since $\|g_0({\bf v}_n)-g_0({\bf v}_0)\|= \|(0,g_0({\bf v}_n))-(0,g_0({\bf v}_0))\|=\|(0,g_0({\bf v}_n))-(0,g_0({\bf v}_0))+g(s_n,{\bf w}_n)-g(s_n,{\bf w}_n)\|\leq \|g(s_n,{\bf w}_n)-(0,g_0({\bf v}_n))\|+ \|g(s_n,{\bf w}_n)-(0,g_0({\bf v}_0))\|$, on also has that $g_0({\bf v}_n)\rightarrow g_0({\bf v}_0)$. QED.

\vspace{4mm}
By combining Lemma \ref{l:3} with Lemma \ref{l:4}, and by taking into account that $(f^{-1})\hat{ }=(\hat{f})^{-1}$ and that $(g\cdot f)\hat{ }=\hat{g}\cdot \hat{f}$, one can prove Theorem \ref{t:1}. The details of the proof are omitted.

As in the nonasymptotic case, it is possible for a transformation to be asymptotically regular while its inverse is not. Consider for instance the transformation $f(t,{\bf x}):=(t,{\bf x}/t)$, for which $f^+({\bf v})=0$ holds, and whose inverse $f^{-1}(t,{\bf x})=(t,t{\bf x})$ is not asymptotically regular at any point. It does not appear to be trivial to find a rigorous proof to the following reasonable statement, which is therefore proposed here as a conjecture:

\begin{congettura} \rm
If $f$ is asymptotically regular and $f^+$ is a homeomorphism on $V$, then $f$ is an asymptotic homeomorphism.
\end{congettura}

Hereafter, unless otherwise stated, the term {\it transformation} will be used to designate a causal transformation which is also an asymptotic homeomorphism.


\subsubsection{Aymptotically Euclidean transformations.}

The following classes of transformations are particularly important:

\begin{definizione} \rm
(a) An {\it asymptotically identical} transformation is a transformation $f$ such that $f^+({\bf v})={\bf v}$.
(b) An {\it asymptotically Euclidean} transformation is a transformation $f$ such that $f^+({\bf v})=R{\bf v}-{\bf v}_0$.
\end{definizione}
\begin{lemma} \label{l:5} \rm
(a) Asymptotically identical transformations and asymptotically Euclidean transformations are groups.
(b) An asymptotically Euclidean transformation $f$ can univocally be factorized as $f_1\cdot f_2\cdot f_3$, where $f_3$ is a rotation, $f_2$ is asymptotically identical and $f_1$ is a boost.
(c) If $\gamma _1$ and $\gamma _2$ are two semitrajectories, the two following statements are equivalent: (i) $\omega (\gamma _1)=\omega (\gamma _2)$; (ii) there exists an asymptotically identical transformation $f$ such that $\gamma _2=f(\gamma _1)$.
\end{lemma}


\subsubsection{N-bigbang.}

Let us now consider the case of $N$ semitrajectories, whose particularities with respect to the case of a single semitrajectory it is convenient to note. Furthermore, for reasons that will become clear later, let us require that the N semitrajectories share a common origin. So let us give the following definition:

\begin{definizione} \rm
N-bigbang is the union of $N$ semitrajectories that are disjoined everywhere but in the origin, which is common.
\end{definizione}

The term N-bigbang derives from the fact that, at the origin, all the particles are concentrated at a single point. The reason for requiring the semitrajectories to be disjoined everywhere else is that this allows to demonstrate the following Theorem \ref{t:2}. All the transformations on $G$ preserve the N-bigbangs, since they can neither separate the semitrajectories at the origin nor transform disjoined trajectories into intersecting trajectories. The notion of N-bigbang is relevant to this paper because it will represent the ideal model of universe that will be proposed.

Let $B$ be the set of N-bigbangs. The asymptotic velocity of an N-bigbang is a vector belonging to $V^N$. We again use the symbol $\omega$ to indicate the application $\omega :B \rightarrow V^N$ which associates with every N-bigbang its asymptotic velocity. If $g$ is a transformation on $V$ and $v=({\bf v_1},...,{\bf v}_N)\in V^N$, then $g(v)$ is the vector $(g({\bf v}_1),...,g({\bf v}_N))$; if ${\bf v}_0\in V$, then $Rv-{\bf v}_0$ is the vector $(R{\bf v}_1-{\bf v}_0,...,R{\bf v}_N-{\bf v}_0)$.

The following theorem that holds for N-bigbangs is analogous to Lemma \ref{l:5}c for semitrajectories.

\begin{teorema} \label{t:2} \rm
Let $\beta _1$ and $\beta _2$ be two N-bigbangs. The following two statements are equivalent: (i) $\omega (\beta _1)=\omega (\beta _2)$; (ii) There exists an asymptotically identical transformation $f$ such that $\beta _2=f(\beta _1)$.
\end{teorema}
{\bf Proof.} The implication (ii)$\Rightarrow$ (i) is obvious. An intuitive proof of the implication (i)$\Rightarrow$ (ii) is given. By using the induction principle, let us suppose that the theorem holds for an N-bigbang and prove that it also holds for an (N+1)-bigbang. Let $\beta$ and $\beta '$ be two (N+1)-bigbangs for which the equation $\omega (\beta )=\omega (\beta ')$ holds. Due to our hypothesis of induction, there exists a transformation $f$ such that $\gamma '_i=f(\gamma_i)$, for $i=1,...,N$, where $\gamma_i$ and $\gamma'_i$ are the semitrajectories that form the (N+1)-bigbangs. It is therefore sufficient to prove that given an N-bigbang $\beta$ and two semitrajectories $\gamma$  and $\gamma '$ such that $\omega (\gamma )=\omega (\gamma ')$, there exists an asymptotically identical transformation $f$ such that $f(\beta )=\beta$  and $f(\gamma )=\gamma '$.

To intuitively accept the existence of such a transformation, consider two points $\gamma (t)$ and $\gamma '(t)$ which evolve in three-dimensional space along with the segment that joins them. If the segment intersects one of the other points of the N-bigbang, it is allowed to fold slightly in order to avoid it (note that in two-dimensional space this is not possible). Therefore, there exists at every given instant a broken line which joins the two points and whose length is less than $2\|\gamma (t)-\gamma '(t)\|$. Consider now the ``tube'' of radius $\epsilon$  that surrounds the segment, i.e. the set of points whose distance from the segment is less than $\epsilon$ . At every instant it is possible to determine $\epsilon$  so that the tube does not intersect any of the remaining $N$ points. Finally, consider for every $t$ a space transformation $f_X(t,\cdot)$ which keeps the space outside the tube unmodified and deforms it inside the tube, bringing the point $\gamma (t)$ onto the point $\gamma '(t)$. The space-time transformation required will therefore be $f(t,{\bf x}):=(t,f_X(t, {\bf x}))$. Since for every $t$ the relation $\|f_X(t, {\bf x})-{\bf x}\|\leq 2\left(\|\gamma (t)-\gamma '(t)\|+\epsilon\right)$ holds, one can easily prove that $f$ is asymptotically identical. QED.


\subsubsection{Asymptotic intervals.} \label{s216}

This section proves a theorem and a corollary which will be subsequently used to demonstrate the invariance of the asymptotic quantum measure under AET.

First of all, let us introduce some notations: if $\Delta \subseteq V$, let us define $\Delta^c:=\{(t, {\bf v}t)\in G|t\geq 0, {\bf v}\in \Delta \}$; in particular, if ${\bf v}\in V$, then ${\bf v}^c:=\{{\bf v}\}^c$ is the semitrajectory $\{{\bf v}t\}_{t\geq0}$. One can easily see that if $f$ is asymptotically identical, then $\|f({\bf v}^c)(t)-{\bf v}^c(t)\|/t\rightarrow 0$ for $t\rightarrow +\infty$  (one should bear in mind that an asymptotically identical transformation preserves the asymptotic limit of semitrajectories). The following theorem extends this property:

\begin{teorema} \label{t:3} \rm
Let $\Delta \subseteq V$ be a bounded set and let $f$ be an asymptotically identical transformation. Then $\sup_{{\bf v}\in \Delta} \|f({\bf v}^c)(t)-{\bf v}^c(t)\|/t\rightarrow 0$ for $t\rightarrow +\infty$.
\end{teorema}
{\bf Proof}. In this case as well, for the sake of simplicity, let us think of $f$ as a causal transformation on $G^+$. Since $\sup_{{\bf v}\in \Delta} \|f({\bf v}^c)(t)-{\bf v}^c(t)\|\leq \sup_{{\bf v}\in \bar{\Delta}} \|f({\bf v}^c)(t)-{\bf v}^c(t)\|$, we can directly assume that $\Delta$  is compact. Let us set $d(t):=\|f({\bf v}^c)(t)-{\bf v}^c(t)\|/t$. Obviously, $d(t)\rightarrow 0$ if and only if $d(f_T(t))\rightarrow 0$. Due to Lemma \ref{l:2}d, one finds that $d(f_T(t))= \sup_{{\bf v}\in \Delta} \|f_X(t,{\bf v}t)/f_T(t)-{\bf v}\|$. Since $\hat{f}(1/t,{\bf v})=(1/f_T(t),f_X(t,{\bf v}t)/f_T(t))$ (where $\hat{f}$ is defined in section \ref{s213}), therefore $d(f_T(t))=\sup_{{\bf v}\in \Delta} \|g_{(_1/t)}({\bf v}) -{\bf v}\|$, where $g_s({\bf v}):=P_V[\hat{f}(s,{\bf v})]$. Due to the well known property of uniformly convergent functions, if one proves that for $s\rightarrow 0$ the function $g_s({\bf v})$ uniformly tends to ${\bf v}$ in $\Delta$ , the theorem is proven.

In order to prove the uniform convergence of $g_s({\bf v})$, we use the fact that a continuous function on a compact set is also uniformly continuous. From Lemma \ref{l:4}a, the function
$$ \label{02}
\bar{g}(s,{\bf v}):=\cases{\hat{f}(s,{\bf v})$ for $s>0 \cr
(0,{\bf v})$ for $s=0}
$$
is continuous on $F_0^+$, and therefore uniformly continuous on $[0,c]\times\Delta \subseteq F_0^+$, where $c$ is any constant greater than $0$. Hence, for any fixed $\epsilon$, there exists a $\delta$ such that if $(s_1,{\bf v}_1),(s_2,{\bf v}_2)\in[0,c]\times\Delta$ and $\|(s_1,{\bf v}_1)-(s_2,{\bf v}_2)\|<\delta$ then $\|\bar{g}(s_1,{\bf v}_1)- \bar{g}(s_2,{\bf v}_2)\|<\epsilon$. In particular, if one chooses $s_2=0$ and ${\bf v}_1={\bf v}_2={\bf v}$, if $\|(s,{\bf v})-(0,{\bf v})\|=s<\delta$, then $\|\bar{g}(s,{\bf v})-(0,{\bf v})\|<\epsilon$, $\forall {\bf v}\in \Delta$. Since $\|g_s({\bf v})-{\bf v}\|\leq \|\bar{g}(s,{\bf v})-(0,{\bf v})\|$, the theorem is proven. QED.

\vspace{4mm}
Given ${\bf a},{\bf b}\in V$, with $a_i<b_i$, $i=1,2,3$, let $I:=( {\bf a},{\bf b}]=(a_1,b_1]\times(a_2,b_2]\times(a_3,b_3]$ be a half-open interval of $V$; furthermore, if $0<\epsilon <\min\{(b_1-a_1),(b_2-a_2),(b_3-a_3)\}/2$, let $I_{-\epsilon}$ and $I_{+\epsilon}$ be the intervals $({\bf a}+\epsilon , {\bf b}-\epsilon ]$ and $({\bf a}-\epsilon , {\bf b}+\epsilon ]$, respectively.

\begin{corollario} \label{c:1} \rm
Let $f$ be asymptotically identical and let $I=({\bf a},{\bf b}]$ be an interval of $V$. For every $\epsilon$ such that $0<\epsilon <\min\{(b_1-a_1),(b_2-a_2),(b_3-a_3)\}/2$, there exists a $t_0$ such that if $t\geq t_0$, then $I_{-\epsilon}^c(t)\subseteq f(I^c)(t)\subseteq I_{+\epsilon}^c(t)$.
\end{corollario}
{\bf Proof}. From Theorem \ref{t:3}, for every $\epsilon >0$ there exists a $t_0$ such that for $t\geq t_0$ the relation $\sup_{{\bf v}\in I}\|f({\bf v}^c)(t)-{\bf v}^c(t)\|\leq \epsilon$ holds. This means that $I_{-\epsilon}^c(t)\subseteq f(I^c)(t)\subseteq I_{+\epsilon}^c(t)$. QED.

\vspace{4mm}
The above corollary can easily be generalized to the intervals of $V^N$.


\subsection{Classical Trajectories.} \label{s22}

Suppose that $N$ masses ${m_1,...,m_N}$ and $N(N-1)/2$ potentials $V_{i,j}(r)=V_{j,i}(r)$, $i,j\in {1,...,N}$ are given which we assume to be bounded, differentiable and vanishing for $r\rightarrow \infty$ with standard  conditions (which include long-range potentials such as the Coulomb potential; see \cite{scattering} for details).

\begin{definizione} \rm
An N-bigbang (not necessarily an asymptotically regular one) is {\it classical} if it satisfies the equations of motion deriving from the Lagrangian
\begin{equation} \label{03}
L(t)=\sum_{i=1}^N\frac{1}{2}m_i\dot{\gamma}_i^2(t)+\sum_{i<j}V_{i,j}\left(\|\gamma_i(t)-\gamma_j(t)\|\right),
\end{equation}
where $\gamma_i$ are the semitrajectories that make up the N-bigbang.
\end{definizione}

Let $B_C$ be the set of the classical N-bigbangs. Obviously, the set $B_C$ is invariant under Galilean transformations. The following theorem is derived from an important theorem of classical scattering theory, which states that classical trajectories are asymptotically regular:

\begin{teorema} \label{t:4} \rm
Classical N-bigbangs are asymptotically regular.
\end{teorema}
{\bf Proof}. See \cite{scattering}.

\vspace{4mm}
In physical terms, particles asymptotically tend to separate into a certain number of independent clusters. The velocities of the centers of mass of the clusters tend to a constant value. The positions of the particles within each individual cluster with respect to the center of mass of the cluster are bounded or almost bounded, and in any case their asymptotic velocity is the limit velocity of the cluster's center of mass.

Let us set $\Delta_C:=\omega(B_C)\subseteq V^N$. The set $\Delta_C$ is invariant under Euclidean transformations on $V^N$. In general, the equation $\Delta _C=V^N$ does not hold (see next section).

It is useful to define on $B_C$ the following equivalence relations:

\begin{definizione} \rm
 (a) Two classical N-bigbangs $\beta _1$ and $\beta _2$ are said to be {\it G-equivalent} if there exists on $G$ a Galilean transformation $f$ such that $\beta _2=f(\beta _1)$.
 (b) Two classical N-bigbangs $\beta _1$ and $\beta _2$ are said to be {\it $\omega E$-equivalent} if there exists on $V^N$ a Euclidean transformation $g$ such that $\omega (\beta _2)=g[\omega (\beta _1)]$.
\end{definizione}

In general, there can be $\omega E$-equivalent N-bigbangs which are not $G$-equivalent. In particular, there can be N-bigbangs with the same asymptotic velocity which are not connected by a translation. This can be easily understood if one thinks of asymptotic velocity as the limit of the proper boundary condition of Hamilton's action principle (see next section). The operation of taking the limit for $t\rightarrow +\infty$ induces a further degeneration, due to the fact that all the bound states of a cluster merge into the same asymptotic velocity.


\subsubsection{Asymptotic velocities and asymptotic boundary conditions.} \label{s221}

Consider the classical N-bigbangs with origin at the point $(0,0)$ and traveling through the point $(t,x), \, x\in R^{3N}$. These are the proper boundary conditions of Hamilton's action principle, and they are known to generally define more than one trajectory. Suppose we set $x=vt$ and take $t$ to infinity. We will say that the velocity $v$ was set as {\it asymptotic boundary condition} for the N-bigbang. The following questions naturally arise: In which domain $\Delta_B\subseteq V^N$ are the asymptotic boundary conditions defined, i.e., for which values of $v$ can one take the limit? Is the asymptotic velocity of the N-bigbangs obtained in this way equal to $v$? What happens if $v\not\!\!{\in}\Delta_C$? The following simple example helps to show what might happen.

Let us consider, for the sake of simplicity, a one-dimensional particle of mass m subject to the potential
\begin{equation} \label{04}
V(x)=\cases{V_0>0$ for $|x|\leq a \cr 0$ for $r>a \cr}.
\end{equation}
The semitrajectory $x(t)$ that starts at $(0,0)$ with the initial velocity $v_I>0$ is
\begin{equation} \label{05} 
x(t)=\cases{v_It$ for $t\leq a/v_I \cr a+(t-a/v_I)\sqrt{v_I^2+2V_0/m}$ for $t>a/v_I \cr}.
\end{equation}
An analogous equation defines the trajectory with $v_I<0$, while if $v_I=0$ the trajectory is simply $x(t)=0$. One can easily see that $\Delta _C=\left(-\infty ,-\sqrt{2V_0/m}\right)\cup \{0\}\cup \left(\sqrt{2V_0/m},+\infty\right)$.

Let us now set the boundary condition $x(t)=vt$, then we will take $t$ to infinity and we will compute $v_I$ as a function of $v$. If $v>0$, by setting the boundary condition $x(t)=vt$, for large enough values of $t$ one obtains from (\ref{05}) the equation
\begin{equation} \label{06}
a+(t-a/v_I)\sqrt{v_I^2+2V_0/m}=vt,
\end{equation}
which is a fourth-degree equation in $v_I$. One can easily obtain a solution for $t\rightarrow +\infty$ if one assumes that $v_I(t)\rightarrow v_{I\infty}$ , i.e., if one allows that $v_I(t)$ admits a limit, for $t\rightarrow +\infty$. One has to distinguish between two cases: $v_{I\infty} \neq 0$ and $v_{I\infty} =0$. In the former, for very large values of $t$ one can disregard $a$ and $a/v_I$ as compared to $t$ in expression (\ref{06}), and obtain
$$ \label{07}
\sqrt{v_I^2+2V_0/m}=v,
$$
hence
\begin{equation} \label{08}
v_{I\infty}=\sqrt{v^2-2V_0/m}.
\end{equation}
This solution is acceptable only if $v>\sqrt{2V_0/m}$. By replacing $v_{I\infty}$  in expression (\ref{05}), one finds that the asymptotic velocity of the trajectory is $v$ and is therefore equal to the asymptotic boundary condition.

In the case $v_{I\infty}=0$, again for very large values of $t$, one can disregard the term $v_I^2$ with respect to the term $2V_0/m$ in expression (\ref{06}). One obtains
$$ \label{09}
a+(t-a/v_I)\sqrt{2V_0/m}=vt,
$$
hence
\begin{equation} \label{10}
v_I=\frac{a\sqrt{2V_0/m}}{t\left(\sqrt{2V_0/m}-v\right)+a}.
\end{equation}
This solution, which holds only for $0\leq v<\sqrt{2V_0/m}$ because $v_I$ must be $\geq 0$, confirms the correctness of the Ansatz $v_{I\infty}=0$. The case $v=\sqrt{2V_0/m}$ is a limit case, and by straightforward reasoning one can deduce that it forces $v_{I\infty}$ to be null. In a similar way one can calculate the dependence of $v_{I\infty}$ upon $v$ in the case of $v<0$, while $v=0$ trivially implies $v_{I\infty} =0$.

In conclusion, one can say that in this example any value of the velocity $v$ is allowed as asymptotic boundary condition; therefore the set of asymptotic boundary conditions $\Delta_B$ is different and wider than the set of asymptotic velocities $\Delta_C$. Any value $|v|>\sqrt{2V_0/m}$ determines a trajectory with asymptotic velocity $v$, while all the values $|v|\leq\sqrt{2V_0/m}$  determine the same trajectory with null initial velocity and null asymptotic velocity. One could therefore say that for $v_I=0$ there is a ``degeneration'' of the asymptotic boundary conditions. Notice that the point $v_I=0$ is a discontinuity point of the application $\omega _V:V_I\rightarrow V$ that associates the initial velocity with the asymptotic velocity:
\begin{equation} \label{11}
\omega_V(v_I)=\cases{$sign $(v_I)(|v_I|+\sqrt{2V_0/m})$ for $v_I\neq 0 \cr
0$ if $v_I=0 \cr}.
\end{equation}

This simple example gives a glimpse of the possibility of developing an interesting theory on asymptotic boundary conditions. In particular, this theory should: (i) Provide a definition of asymptotic boundary conditions which is more rigorous and general than the one provided above, which is likely to become inadequate in the presence of bound states; (ii) Define the existence domain $\Delta_B$ of the asymptotic boundary conditions \footnote{One thing that can be easily said about $\Delta_B$ is that it is invariant for Euclidean transformations on $V^N$.} ; (iii) Describe the relation between asymptotic velocity and asymptotic boundary condition, proving for example that if the asymptotic boundary condition belongs to $\Delta _C$, then it is equal to the asymptotic velocity; (iv) Describe the relation between degeneration points of asymptotic boundary conditions and discontinuity points of the function $\omega_V$.

I am not aware of the existence of such a theory, and it will not be developed in this paper.


\subsection{Asymptotic quantum measure.}

This section is devoted to quantum mechanics. The first part defines and studies quantum asymptotic velocity. This operator will be used for the definition of the asymptotic quantum measure in the second part of this section. The third part will prove the invariance of asymptotic quantum measure under AET, which is probably the main mathematical result of this paper.

Consider an N-particle quantum system. Its state is described by the vector $\psi \in L^2(R^{3N})$; $Q=({\bf Q}_1,..., {\bf Q}_N) =(Q_{1x},Q_{1y},Q_{1z},...,Q_{Nx},Q_{Ny},Q_{Nz})$ is the position vector operator and $P$ (of analogous structure) is the momentum operator. The Hamiltonian is
\begin{equation} \label{12}
H=\sum_{i=1}^N\frac{{\bf P}_i^2}{2m_i}+\sum_{i<j}V_{i,j}(\|{\bf Q}_i-{\bf Q}_j\|),
\end{equation}
where the same considerations of section \ref{s22} apply to the potentials $V_{i,j} $.

\subsubsection{Quantum asymptotic velocity}

\begin{definizione} \rm
We call {\it quantum asymptotic velocity} the vector operator
\begin{equation} \label{13}
V^+:=s-\lim_{t\rightarrow+\infty}e^{iHt}\frac{Q}{t}e^{-iHt}.
\end{equation}
\end{definizione}

Quantum asymptotic velocity also is an important concept used in quantum scattering theory, and its existence is assured by a theorem which is analogous to the classical case:

\begin{teorema} \label{t:5} \rm The limit (\ref{13}) exists for a dense subset of  $L^2(R^{3N})$. $V^+$ is a vector of Hermitian operators which commute with one another and with the Hamiltonian.
\end{teorema}
{\bf Proof}. See \cite{scattering}.
\vspace{4mm}

Let us give the explicit expression for $V^+$ in two simple cases. For a free particle one has:
\begin{equation} \label{14}
{\bf V}^+=\frac{{\bf P}}{m}.
\end{equation}
One can easily obtain equation (\ref{14}) by using the relation $e^ABe^{-A}=B+[A,B]+\frac{1}{2!}[A,[A,B]]+...$. The physical interpretation of this result is simple: for very large times, measuring the position of a particle at the time $t$ and dividing it by $t$ is equivalent to measuring its velocity.
For a particle in a central potential which admits the M\o ller operators $\Omega ^\pm$, one has
\begin{equation} \label{15}
{\bf V}^+=\Omega^{-}\frac{\bf P}{m}\Omega^{-\dagger}E_S,
\end{equation}
where $E_S$ is the projector over the scattering states of the Hamiltonian. See \cite{scattering} for the proof. Note that the bounded states of the Hamiltonian belong to the eigenvalue $0$ of ${\bf V}^+$. This result holds in general.

We derive now the transformation rules for asymptotic velocity. First of all, let us summarize the Galilean transformations and their generators for an N-particle quantum system: (i) Time translations: $e^{-itH}$. (ii) Space translations: $e^{-i{\bf P}\cdot {\bf x}}$, where ${\bf P}=\sum_i {\bf P}_i$ is the total momentum; one has: $e^{i{\bf P} \cdot {\bf x}}Qe^{-i{\bf P} \cdot {\bf x}}=Q+{\bf x}$. The momentum and the Hamiltonian are invariant under space translations. (iii) Rotations: $e^{-i{\bf L}\cdot{\bf \alpha}}$, where ${\bf L}=\sum_i{\bf L}_i$ is the total angular momentum and ${\bf \alpha}$ is the vector associated with a rotation; one has $e^{i{\bf L}\cdot {\bf \alpha}}Qe^{-i{\bf L}\cdot {\bf \alpha}}=R_{\bf \alpha} Q$, where $R_{\bf \alpha}$  is the orthogonal matrix associated with the vector ${\bf \alpha}$ . The Hamiltonian is invariant under rotations. (iv) Boosts: $e^{-im{\bf Q}\cdot {\bf v}}$, where $m{\bf Q}=\sum_im_i{\bf Q}_i$; we have $e^{im{\bf Q}\cdot {\bf v}}{\bf P}_ie^{-im{\bf Q}\cdot {\bf v}}={\bf P}_i-m_i{\bf v}$ and $e^{im{\bf Q}\cdot {\bf v}}He^{-im{\bf Q}\cdot {\bf v}}= H+{\bf v}\cdot {\bf P}+\frac{1}{2}{\bf v}^2\sum_i m_i$.
\begin{lemma} \label{l:6} \rm The operators $V^+$ have the following transformation rules:
\begin{eqnarray} \label{ar1}
e^{iHt}V^+e^{-iHt} & = & V^+, \nonumber \\
e^{i{\bf P}\cdot {\bf x}}V^+e^{-i{\bf P}\cdot {\bf x}} & = & V^+, \\
e^{i{\bf L}\cdot{\bf \alpha}}V^+e^{-i{\bf L}\cdot{\bf \alpha}} & = & R_{\bf \alpha} V^+, \nonumber \\
e^{im{\bf Q}\cdot{\bf v}}V^+e^{-im{\bf Q}\cdot {\bf v}} & = & V^+-{\bf v}. \nonumber
\end{eqnarray}
\end{lemma}
{\bf Proof}. The proof can be obtained by simple calculations based on the transformation rules of $Q$, $P$, and $H$, and taking into account that $V^+$ commutes with the Hamiltonian.


\subsubsection{Asymptotic quantum measure.} \label{s232}

Let $E^+(\Delta )$ be the spectral measure on $V^N$ associated with the operators $V^+$, where $\Delta$  belongs to the Borel $\sigma$-algebra ${\cal B}$ of $V^N$. Choosing a state $\psi$, one can induce on $V^N$ a measure $\mu$  by defining $\mu (\Delta ):= \langle \psi |E^+(\Delta )|\psi \rangle$. The state we choose to define the quantum measure $\mu_Q$ is the following: if ${\bf x}\in R^3$, ${\bf x}_N$ is the vector $({\bf x}_1,..., {\bf x}_N)\in R^{3N}$ such that ${\bf x}_i={\bf x},\, i=1,...,N$. The vector ${\bf x}_N$ describes all the particles concentrated at the point ${\bf x}$. The state $|{\bf x}_N\rangle$  is the improper eigenvector with eigenvalue ${\bf x}_N$ of the position operator $Q$.

\begin{definizione} \rm
The {\it asymptotic quantum measure} (or more simply {\it quantum measure}) is the measure $\mu_Q$ on $V^N$ defined as
\begin{equation} \label{16}
\mu_Q(\Delta ):=\langle {\bf x}_N|E^+(\Delta )| {\bf x}_N\rangle.
\end{equation}
\end{definizione}
{\bf Example}. One can easily calculate $\mu_Q$ when there is no potential and the equation $V^+=({\bf P}_1/m_1,..., {\bf P}_N/m_N)$ holds. If $E^P(\cdot )$ is the spectral measure of $P$, then $E^+(\Delta )=E^P(m\Delta )$, where $m\Delta := \{(m_1{\bf v}_1,...,m_N{\bf v}_N)\in R^{3N}|({\bf v}_1,...,{\bf v}_N)\in \Delta \}$. Therefore
\begin{eqnarray} \label{ar2}
\mu_Q(\Delta) & = &\langle {\bf x}_N|E^+(\Delta )| {\bf x}_N\rangle=
\int{\langle {\bf x}_N|p_1\rangle dp_1\langle p_1|E^P(m\Delta)|p_2\rangle dp_2\langle p_2|{\bf x}_N\rangle}= \nonumber \\
& = & \frac{1}{(2\pi)^{3N}}\int{\exp[-i{\bf x}_N(p_2-p_2)]\chi_{m\Delta}(p_1)\delta(p_2-p_1)dp_1dp_2}= \\
& = & \frac{1}{(2\pi)^{3N}}\int{\chi_{m\Delta}(p)dp}=\frac{\mu_L(m\Delta)}{(2\pi)^{3N}} \nonumber,
\end{eqnarray}
where $\chi_{m\Delta}$ is the characteristic function of the set $m\Delta$  and $\mu_L$ is the Lebesgue measure. The measure $\mu_Q$ in the absence of a potential is therefore proportional to the Lebesgue measure on the momentum space.
\vspace{4mm}

The following lemma describe the invariance properties of the measure $\mu_Q$:

\begin{lemma} \label{l:7} \rm
(a) $\mu_Q$ does not depend on the point $\,{\bf x}$. (b) $\mu_Q$ is invariant under Euclidean transformations on $V^N$, i.e., $\mu_Q(\{R\Delta -{\bf v}_0\})=\mu_Q(\Delta )$, where $\{R\Delta -{\bf v}_0\}:=\{(Rv-{\bf v}_0)\in V^N|v\in \Delta \}$.
\end{lemma}
{\bf Proof}. Here also the proof can be obtained by simple calculations based on the transformation rules of $V^+$.
\vspace{4mm}

The invariance of the measure $\mu_Q$ under Euclidean transformations allows to define a quotient measure $\tilde{\mu}_Q$ on the quotient $\sigma$-algebra $\tilde{\cal B}$, which is formed by the measurable sets of $V^N$ which are also invariant under Euclidean transformations. See the Appendix.


\subsubsection{Invariance of the quantum measure for asymptotically Euclidean transformations.}

\begin{lemma} \label{l:8} \rm
The equation
\begin{equation} \label{17}
\mu_Q(\Delta)=\lim_{t\rightarrow +\infty}\int\limits_{\Delta^c(t)}{|K({\bf x}_N,y,t)|^2\,{\rm d}y}
\end{equation}
holds, where $K({\bf x}_N,y,t)$ is the Feynman propagator between points $(0,{\bf x}_N)$ and $(t,x)$, and where $\Delta^c$ has been defined in section \ref{s216}.
\end{lemma}
{\bf Proof}. If $E^Q_x$ is used to indicate the spectral family of $Q$, the equation
$$ \label{18}
e^{iHt}\frac{Q}{t}e^{-iHt}=e^{iHt}\frac{1}{t}\left[\int{x\,dE_x^Q}\right]e^{-iHt}=
\int{v\,d(e^{iHt}E_{vt}^Qe^{-iHt})}
$$
allows one to deduce that, for the spectral family $E^+_v$ of $V^+$, the equation
\begin{equation} \label{19}
E_v^+=s-\lim_{t\rightarrow+\infty}e^{iHt}E_{vt}^Qe^{-iHt}
\end{equation}
holds, and therefore for the spectral measure $E^+(\Delta )$ one has that
\begin{equation} \label{20}
E^+(\Delta)=s-\lim_{t\rightarrow+\infty}e^{iHt}E^Q[\Delta^c(t)]e^{-iHt}
\end{equation}
holds.
Equation (\ref{17}) can be easily derived from equations (\ref{16}) and (\ref{20}). QED.
\vspace{4mm}

By expressing the Feynman propagator in terms of a sum over paths, one has
\begin{equation} \label{21}
\mu_Q(\Delta)=\lim_{t\rightarrow +\infty}\int\limits_{\Delta^c(t)}{{\rm d}y\int\limits_{(0,{\bf x}_N)}^{(t,y)}{{\cal D}\beta_1{\cal D}\beta_2 \, S(\beta_1)S^*(\beta_2)}},
\end{equation}
where $\beta _1$ and $\beta _2$ are two paths with extremes $(0,{\bf x}_N)$ and $(t,y)$, and
$$ \label{22}
S(\beta_i):=\exp\left[-i\int_{0}^{t}{L\left(\beta_i(\tau),\dot{\beta}_i(\tau)\right)\,{\rm d}\tau}\right],
$$
where $L$ is the Lagrangian of the system. The following theorem establishes the invariance of the quantum measure under AET:
\begin{teorema} \label{t:6} \rm
If $f$ is an asymptotically Euclidean transformation, then the equation
\begin{equation} \label{23}
\mu_Q(\Delta)=\lim_{t\rightarrow +\infty}\int\limits_{\Delta^c(t)}{{\rm d}y\int\limits_{(0,{\bf x}_N)}^{(t,y)}{ {\cal D}\beta_1{\cal D}\beta_2 \, S[f(\beta_1)]S^*[f(\beta_2)]}}
\end{equation}
holds.
\end{teorema}
{\bf Proof}. Since we have already proved the invariance of $\mu_Q$ under translations of the point ${\bf x}$ and under Galilean transformations, for the sake of simplicity we demonstrate equation (\ref{23}) only in the case of a transformation $f$ which is asymptotically identical and leaves the point $(0,{\bf x})$ unmodified. One has
\begin{eqnarray} \label{ar3}
\int\limits_{\Delta^c(t)}{{\rm d}y\int\limits_{(0,{\bf x}_N)}^{(t,y)}{ {\cal D}\beta_1{\cal D}\beta_2 \, S[f(\beta_1)]S^*[f(\beta_2)]}}= \nonumber \\
\int\limits_{f_X[t,\Delta^c(t)]}{{\rm d}y\int\limits_{(0,{\bf x}_N)}^{(f_T(t),y)}{ {\cal D}\beta_1{\cal D}\beta_2 \, S(\beta_1)S^*( \beta_2)}}= \\
\int\limits_{f(\Delta^c)[f_T(t)]}{|K({\bf x}_N,y,f_T(t))|^2\,{\rm d}y}, \nonumber
\end{eqnarray}
where $f_T$ and $f_X$ were defined in section \ref{s212}, and Lemma \ref{l:2}d has been used. If we disregard border effects, which as we will see are nullified by the condition of asymptotic identity, equations (\ref{ar3}) point out the invariance of the Feynman path integrals under a generic causal homeomorphism.

Since $f_T(t)$ is monotonically increasing, in order to prove equation (\ref{23}) is enough to prove that
$$ \label{24}
\lim_{t\rightarrow +\infty}\int\limits_{f(\Delta^c)(t)}{|K({\bf x}_N,y,t)|^2{\rm d}y}=
\lim_{t\rightarrow +\infty}\int\limits_{\Delta^c(t)}{|K({\bf x}_N,y,t)|^2{\rm d}y}.
$$
Let $I$ be a half-open interval of $V^N$, as defined in section \ref{s216}. For the sake of brevity, let
$$ \label{25}
\mu_{Qt}(I):=\int\limits_{I^c(t)}{|K({\bf x}_N,y,t)|^2\,{\rm d}y}.
$$
Owing to Corollary \ref{c:1}, one has, for large enough values of $t$,
$$ \label{26}
\mu_{Qt}(I_{-\epsilon})\leq\mu_{Qt}[f(I)]\leq\mu_{Qt}(I_{+\epsilon});
$$
if one takes the limit for $t\rightarrow +\infty$,
\begin{equation} \label{27}
\mu_Q(I_{-\epsilon})\leq\liminf_{t\rightarrow +\infty}\mu_{Qt}[f(I)]\leq\limsup_{t\rightarrow +\infty}\mu_{Qt}[f(I)]\leq \mu_Q(I_{+\epsilon})
\end{equation}
Furthermore, one has:
\begin{equation} \label{28}
\lim_{\epsilon\rightarrow 0}\left[\mu_Q(I_{+\epsilon})-\mu_Q(I_{-\epsilon})\right]=
\lim_{\epsilon\rightarrow 0}\mu_Q\left[I_{+\epsilon}\setminus I_{-\epsilon}\right]=\mu_Q\left[\bigcap_{\epsilon>0}(I_{+\epsilon}\setminus I_{-\epsilon})\right]=\mu_Q(\partial I).
\end{equation}
where $\partial I$ is the boundary of the interval $I$. The Lebesgue measure of $\partial I$ is null, but this does not ensure that the quantum measure is null as well, because it could be singular with respect to the Lebesgue measure. However, one can prove that $\mu_Q(\partial I)=0$ if the measure $\mu_Q$ is $\sigma$-finite; we will assume this without proof.

Consider a single side of the interval $I$, for instance $D:=\{a_{1x}\}\times(a_{1y},b_{1y}]\times...\times(a_{Nz},b_{Nz}]$, and suppose that $\mu_Q(D)>0$. The set $D_v:=\{D-(v,0,0)\}$ is obtained from set $D$ by means of a translation along the x-axis; therefore, due to the invariance of $\mu_Q$, one has that $\mu_Q(D)=\mu_Q(D_v)$. If ${(v_n,0,0)}$ is a bounded sequence of elements of $V$ with distinct values, one has that $\cup_nD_{v_n}$ is a bounded set; however, since all the $D_{v_n}$ are disjoined, one has that $\mu_Q(\cup_n D_{v_n})=\sum_n\mu_Q(D_{v_n})=+\infty$ , which is in contrast with the hypothesis that $\mu_Q$ is $\sigma$-finite.

If $\mu_Q(\partial I)=0$, then from relation (\ref{28}), and since $I_{-\epsilon} \subseteq I\subseteq I_{+\epsilon}$ , one obtains 
$$ \label{29}
\lim_{\epsilon \rightarrow 0}\mu_Q(I_{+\epsilon} )=\lim_{\epsilon\rightarrow 0}\mu_Q(I_{-\epsilon})= \mu_Q(I),
$$
while from relation (\ref{27}) one obtains
\begin{equation} \label{30}
\lim_{t\rightarrow +\infty}\mu_{Qt}[f(I)]=\mu_Q(I).
\end{equation}

We have demonstrated that equation (\ref{23}) holds when $\Delta$ is an half-open interval of $V^N$. The demonstration is completed by using the fact that according to a well-known theorem of measure theory, if two measures on $R^N$ agree on half-open intervals, then they are equal. QED.


\subsection{Asymptotically Euclidean manifolds.}

The fact that the quantum measure is invariant under AET's means that its definition does not require a space-time provided with metrics, i.e., it demonstrates the {\it non-metric} nature of such a measure. Formulating the AET theory in a coordinate-free context, generated through a mechanism similar to the one used for differential manifolds, points out this property more clearly. This section therefore defines and studies {\it asymptotically Euclidean manifolds.}

\begin{definizione} \rm 
(a) Let $M$ be a set. We call {\it Galilean reference frame} for $M$ a bijective application $\varphi :G\rightarrow M$.
(b) Two Galilean reference frames $\varphi _1$ and $\varphi _2$ are said to be {\it $\omega$E-equivalent} if $\varphi _2^{-1}\cdot\varphi_1:G\rightarrow G$ is an asymptotically Euclidean transformation.
(c) An {\it asymptotically Euclidean manifold} is the pair $(M,{\cal A})$, where ${\cal A}$ is a class of the $\omega E$-equivalence relation.
\end{definizione}

The definitions of asymptotically regular semitrajectories and N-bigbangs, of asymptotic homeomorphism, asymptotically identical transformation and, finally, asymptotically Euclidean transformation can be transferred trivially from $G$ to the manifold $M$ through any one of its reference frames. For instance: a subset $\hat{\gamma} \subseteq M$ is an asymptotically regular semitrajectory on $M$ if $\varphi^{-1} (\hat{\gamma} )$ is an asymptotically regular semitrajectory on $G$ for any reference frame $\varphi \in {\cal A}$. However, the notion of classical N-bigbang cannot be transferred on $M$, because it is not invariant under AET's. In order to indicate the objects on $M$ we use the same symbols used for the corresponding objects on $G$, but writing them with hat. For instance, ${\hat{B}}$ is the set of the asymptotically regular N-bigbangs on $M$.

We will now construct the space of asymptotic velocities for the manifold $(M,{\cal A})$.
\begin{definizione} \rm 
(a) Two semitrajectories $\hat{\gamma} _1$ and $\hat{\gamma} _2$ on $M$ are said to be {\em $\omega$-equivalent} if, for any reference frame $\varphi$, one has $\omega [\varphi^{-1}(\hat{\gamma}_1)]=\omega [\varphi^{-1} (\hat{\gamma} _2)]$. This definition does not depend on the chosen reference frame.
(b) The {\it space of asymptotic velocities on M}, indicated by $\hat{V}$, is the quotient space of the $\omega$-equivalence relation.
\end{definizione}
The space $\hat {V}$ can be considered the asymptotic correspondent of the tangent vector space of differential manifolds. Let us construct the class of reference frames for $\hat{V}$. Let $\hat{\omega}:{\hat{\Gamma}}\rightarrow \hat{V}$ be the application that associates each semitrajectory wich its $\omega$-equivalence class. A reference frame $\varphi$ of $M$ induces an application $\phi_\varphi : V\rightarrow \hat{V}$ which is defined by setting $\phi_\varphi ({\bf v}):=\hat{\omega}[\varphi(\gamma)]$, where $\gamma$ is any semitrajectory such that $\omega(\gamma)={\bf v}$. Let us set ${\cal A}_V:=\{\phi_\varphi |\varphi \in {\cal A}\}$.

\begin{lemma} \label{l:9} \rm 
(a) The applications of ${\cal A}_V$ are bijective.
(b) If $\phi_1,\phi_2\in {\cal A}_V$, then $\phi _2^{-1}\cdot \phi _1$ is a Euclidean transformation on $V$. 
(c) If $\phi \in {\cal A}_V$ and if $g$ is a Euclidean transformation on $V$, then $ \phi\cdot g\in {\cal A}_V$ as well.
\end{lemma}

The set ${\cal A}_V$ is therefore the class of reference frames for $\hat{V}$. Class ${\cal A}_V$ induces on $\hat{V}$ both a structure of affine space, and the definition of Euclidean transformation. The space $\hat{V}^N$ can equally be defined as the Cartesian product of $\hat{V}$ or as the $\omega$-equivalence class of the N-bigbangs of $\hat{B}$. A reference frame for $\hat{V}^N$ is derived trivially from a reference frame for $\hat{V}$. Let $\phi _1$ and $\phi _2$ be two reference frames for $\hat{V}^N$, and let $\Delta \subseteq V^N$ be invariant under Euclidean transformations; since $\phi _2^{-1}\cdot \phi_1$ is a Euclidean transformation on $V^N$, one has that $\phi _1(\Delta )=\phi _2(\Delta )$. This means that an invariant subset of $V^N$ can be transferred without ambiguities onto $\hat{V}^N$. In particular, let us define $\hat{\Delta}_C:=\phi (\Delta _C)$ and $\hat{\Delta}_B:=\phi(\Delta_B)$, where $\Delta_C$ and $\Delta_B$, defined in sections \ref{s22} and \ref{s221}, are, respectively, the set of asymptotic velocities and the set of asymptotic boundary conditions for the classical N-bigbangs. On $\hat{V}^N$ one can define the quantum measure $\hat{\mu}_Q$, setting $\hat{\mu}_Q(\hat{\Delta}):=\mu_Q[\phi^{-1}(\hat{\Delta})]$; in this case also, the definition does not depend on the reference frame chosen.

The definition of a classical reference frame is now given:

\begin{definizione} \rm
(a) Given an N-bigbang $\hat{\beta}$ on M, a reference frame $\varphi$ is said to be a {\it classical reference frame for $\hat{\beta}$} if $\varphi^{-1}(\hat{ \beta} )$ is a classical N-bigbang on $G$. 
(b) The {\it classical image} of an N-bigbang $\hat{\beta}$ is the set $\{\varphi^{-1}(\hat{\beta} )\in B |\varphi$ is a classical reference frame for $\hat{\beta} \}$.
\end{definizione}

\begin{lemma} \label{l:10} \rm
(a) An N-bigbang $\hat{\beta}$  admits a classical reference frame if and only if $\hat{\omega} (\hat{\beta} )\in \hat{\Delta} _C$.
(b) If there exists on $M$ an asymptotically Euclidean transformation $\hat{f}$ such that for two N-bigbangs $\hat{\beta}_1$ and $\hat{\beta}_2$ one has $\hat{\beta}_2=f(\hat{\beta}_1)$, then $\hat{\beta} _1$ and $\hat{\beta} _2$ have the same classical image.
(c) The classical image of an N-bigbang is a class of the $\omega E$-equivalence relation on $B_C$ (defined in section \ref{s22}).
\end{lemma}
{\bf Proof}. Only the proof of point (a) is given. (i) $\hat{\beta}$ admits a classical reference frame $\Rightarrow \hat{\omega} (\hat{\beta} )\in \hat{\Delta}_C$: if $\varphi$ is a classical reference frame for $\hat{\beta}$, the relation $\omega [\varphi^{-1}(\hat{\beta})]\in \Delta_C$ must hold; since $\hat{\omega}(\hat{\beta})=\phi_\varphi[\omega(\varphi^{-1}(\hat{\beta}))]$, the relation $\hat{\omega}(\hat{\beta})\in \hat{\Delta} _C$ also must hold. (ii) $\hat{\omega} (\hat{\beta})\in \hat{\Delta} _C \Rightarrow  \hat{\beta}$  admits a classical reference frame: let $\varphi$ be a generic reference frame; then $\omega [\varphi^{-1} (\hat{\beta} )]\in \Delta _C$. Let $\beta_C\in B_C$ be a classical N-bigbang such that $\omega (\beta_C)=\omega [\varphi^{-1} (\hat{\beta} )]$; by virtue of Theorem \ref{t:2}, there exists on $G$ an asymptotically identical transformation $f$ such that $f(\beta_C)=\varphi^{-1}(\hat{\beta})$. The reference frame $\varphi_C:=\varphi\cdot f$  is therefore a classical reference frame for $\hat{\beta}$. QED.
\vspace{4mm}

One can derive from Lemma \ref{l:10}c that two classical N-bigbangs not connected by a Galilean transformation can belong to the classical image of the same N-bigbang $\hat{\beta}$ on $M$.

The following definition will be useful later:
\begin{definizione} \rm
Let $\hat{\beta}$ be an N-bigbang on $M$. The {\it generalized classical image} of $\hat{\beta}$ is the set of the classical N-bigbangs on $G$ defined by the set of asymptotic boundary conditions $\{\omega [\varphi^{-1}(\hat{\beta})]\in V^N|\varphi \in {\cal A}\}$.
\end{definizione}
Similarly to what happens for the classical image, one can prove that the generalized classical image of an N-bigbang $\hat{\beta}$ is not empty if and only if $\hat{\omega} (\hat{\beta} )\in \hat{\Delta}_B$. The theory mentioned in section \ref{s221} should prove that if $\hat{\omega} (\hat{\beta})\in \hat{\Delta}_B\cap \hat{ \Delta} _C$, its classical image and its generalized classical image coincide. If $\hat{\omega} (\hat{\beta})\in \hat{\Delta}_B\setminus \hat{ \Delta} _C$, then $\hat{\beta}$ admits a generalized classical image but admits no classical reference frame.


\section{Physical interpretation.} 

This second part of the paper discusses the physical and conceptual aspects of the mathematical theory developed in the first part.

\subsection{An ideal model of universe.} \label{s31}

This section proposes an ideal model of universe based on asymptotically Euclidean manifolds. Since this is a nonrelativistic structure, and since the quantum measure is based on bare Schr\"{o}dinger's equation (which is inadeguate to explain many atomic phenomena, see section \ref{s334}), this can only be a schematic and ideal model, and one can draw only interpretative and conceptual conclusions from it. For the sake of brevity, hereafter it will be referenced as {\it asymptotic model}.

The asymptotic model can be described by the following three statements:
\begin{enumerate}
\item The evolution of the particles of the universe is represented by an N-bigbang $\hat{\beta}$  on an asymptotically Euclidean manifold. Two N-bigbangs connected by an asymptotically Euclidean transformation represent the same evolution.
\item (a) The N-bigbang $\hat{\beta}$  appears to us as if it were seen from a classical reference frame. (b) All the classical reference frames for $\hat{\beta}$ are equivalent from an observational standpoint.
\item The quantum measure $\hat{\mu}_Q$ on $\hat{V}^N$ rules the choice of the asymptotic velocity that defines the N-bigbang, in the same way in which, in a probability space, the probability measure rules the choice of an elementary outcome.
\end{enumerate}
Let us examine these three statements. Statements 1 and 2a imply renouncing what we might consider the {\it paradigm} of classical mechanics, i.e., the existence of a metrics defined {\it a priori} on which a law of motion is based which determines the motion of particles starting from the initial conditions \footnote{More in general, one can certainly say that this paradigm is, to a large extent, at the core of our very notion of physical reality.}. Statements 1 and 2a instead state that there is neither a metrics defined a priori nor a law of motion, and that it is the reference frame, i.e., our perception of space distances and time intervals, that structures itself so that particle motion appears to be ruled by a law. It should be noted that Hamilton's action principle already favors an interpretation according to which particle motion is not determined starting from the initial conditions alone, but rather from the initial and final conditions together, which in the asymptotic model are respectively constituted by the initial constrain of the bigbang and by the asymptotic velocity. The asymptotic model takes a further step forward by stating that Hamilton's action principle does not determine the trajectory of the particles but rather the metric. To paraphrase Wheeler et al., who in \cite{wheeler} wrote: ``Time is defined so that motion looks simple'', one could state that: ``{\it Space-time} is defined so that motion looks simple''.

From the observational standpoint, the classical paradigm and this new interpretation are equivalent, since in both cases the particles travel (or appear to travel) along classical trajectories. This new interpretation is proposed here both because it is suggested by the mathematical formalism and because it will make statement 3 more acceptable together with the extremely counterintuitive consequences that it implies, as we will see.

Statement 2a requires the N-bigbang $\hat{\beta}$ to admit a classical reference frame, i.e., it requires its asymptotic velocity to belong to $\hat{\Delta}_C$. Taking into account statement 3, this condition would be certainly verified if the equation
\begin{equation} \label{31}
\mu_Q(V^N\setminus\Delta_C)=0
\end{equation}
were to hold.
The validity of (\ref{31}) obviously depends upon the Lagrangian. Even though the asymptotic model does not provide a Lagrangian, by taking into account the example of section \ref{s221}, condition (\ref{31}) seems exceedingly restrictive. In the mentioned example, the set of asymptotic boundary conditions $\Delta_B$ is wider than the set $\Delta _C$, so that the condition
\begin{equation} \label{32}
\mu_Q(V^N\setminus\Delta_B)=0
\end{equation}
would seem more acceptable [notice that equation (\ref{32}) includes the condition $\Delta_B=V^N$ as a special case]. We will therefore conjecture that for a ``reasonable'' Lagrangian, condition (\ref{32}) should be satisfied. If so, the N-bigbang $\hat{\beta}$  admits a generalized classical image, and it is possible to maintain the logical consistency of the asymptotic model by extending statement 2a in the following way: the N-bigbang $\hat{\beta}$ appears to us as a classical N-bigbang which belongs to its generalized classical image.

Let us move on to statement 2b. Owing to Lemma \ref{l:10}c, this statement can be expressed in an equivalent formulation by saying that two classical N-bigbangs that belong to the same $\omega E$-equivalence class are indistinguishable to our perception. This statement can be broken into the following two statements: (i) two N-bigbangs connected by a Galilean transformation are indistinguishable, and (ii) two N-bigbangs not connected by a Galilean transformation but having the same asymptotic velocity are indistinguishable. Let us examine these two statements separately.

Statement (i) is obvious, since what we perceive are the {\it relative} positions and velocities of particles (which are invariant under Galilean transformations), while we are certainly not able to perceive neither the position and velocity of the center of mass of the universe, nor its space orientation (which are not invariant).

Statement (ii) instead is novel and requires deeper analysis. Consider two different classical N-bigbangs having a common origin, for instance the point $(0,0)$, and having equal asymptotic velocities $v$. The fact that they are indistinguishable could be accounted for physically by stating that they differ only microscopically and therefore that they are {\it macroscopically indistinguishable}. To accept this statement, one should consider that the condition of equal asymptotic velocity can be exceedingly fragile, since it is not met if even a single particle of the universe has a different asymptotic velocity within the two N-bigbangs. For instance, if a photon is emitted in two different space directions (or if it is emitted in one case and not in the other), then the two N-bigbangs will have different asymptotic velocities. It is therefore reasonable to admit that a macroscopic difference between the two N-bigbangs, by irreversibly propagating into the surrounding environment, will lead them to have different asymptotic velocities. Vice versa, two N-bigbangs with the same asymptotic velocity will never be able to differ macroscopically from one another; therefore they will not be distinguishable to our perception. Two N-bigbangs of this kind could correspond, for example, to the two trajectories followed, in the two-slit experiment, by a particle passing through one slit or the other (see section \ref{s332}).

In other words, the macroscopic events of the evolution are determined by the asymptotic boundary conditions of the universe. For instance, two different asymptotic boundary conditions correspond to the fact that I blinked or not while I was writing this sentence. This is not particularly surprising. One should bear in mind that an even stricter rule holds for the {\it initial} conditions (position and velocity): even microscopically different events correspond to different initial conditions of the universe.

Let us return to the general discussion of the asymptotic model. We have seen that according to statement 2a, particles travel (or appear to travel) along classical trajectories. Therefore, in the asymptotic model the wave function is not used to describe particle motion. So how does it explain the quantum phenomena observed in nature? This explanation resides in statement 3 and in the phenomenon of nonchaotic distribution of initial conditions that derives from it. The sections that follow are devoted to the study of these topics.

Let us make a final remark on the big-bang structure of the asymptotic model. Although this type of model for the universe is almost unanimously accepted, its use outside a general-relativistic context might appear simple-minded or strained. However I would like to point out that this structure has played a fundamental role in defining the quantum measure $\mu_Q$ and in proving its invariance under AET's. It is nonetheless reasonable to expect that a much more realistic and elegant big-bang model may be obtained by formulating the theory of asymptotic transformations starting from Einstenian space-time rather from Galilean space-time.


\subsection{The universe as a probability space.}

Statement 3 of the asymptotic model likens the universe to a probability space. A probability space is a tern $(S,{\cal F},\nu )$, where $S$ is the set of elementary outcomes (also called sample space), ${\cal F}$ is the $\sigma$-algebra of the events and $\nu$  is the probability measure. Probability spaces are used in physics to represent a sequence of repetitions of the same experiment, using the following correspondences: the space $S$ corresponds to the possible elementary outcomes of the experiment, the $\sigma$-algebra ${\cal F}$ defines the possible measurable events, and finally the measure of an event, i.e., its probability, corresponds to its relative frequency of occurrence in the sequence.

In this paper, a probability space will be used to represent the universe, employing the following correspondences:
\begin{itemize}
\item The space $S$ is the set $B_C$ of the classical N-bigbangs.
\item The $\sigma$-algebra describes our ability to distinguish the N-bigbangs; therefore its structure must take into account statement 2b of the asymptotic model. This aspect will be considered in detail in section \ref{s322}.
\item The probability measure derives from the quantum measure $\hat{\mu}_Q$. However, the correspondence between probability and relative frequency has to be redefined, since universe evolves only once. This redefinition can be achieved in a simple way, by introducing the concept of {\it sequence of equivalent events} and by making use of the law of large numbers. This new correspondence will be illustrated by a simple example in the next section.
\end{itemize}
The example will furthermore show that not only is it possible to consider the universe as a probability space, but it is {\it necessary} to do so in order to account for the statistical regularities of evolution. The need for a measure in addition to the law of motion to account for the statistical regularities of evolution was noted Popper \cite{popper} and Land\'{e} \cite{lande}, although they used this conclusion as a criticism of determinism.


\subsubsection{A simple example.}

Consider a universe represented by a dynamical system whose phase-space is the interval $[0,1)$ of the real axis, and whose law of motion is Bernoulli's application $f(x):= 2x$ (mod. 1). In this universe, a trajectory is a sequence $\{a_n\}$ of numbers in the interval $[0,1)$ that can be obtained by repeatedly applying the law of motion to an initial condition $a_0$. There is, therefore, a one-to-one correspondence between the trajectories and the initial conditions, i.e., the elements of the interval $[0,1)$. The time of this universe is represented by discrete ticks, which correspond to the natural numbers.

It is very easy to prove that the law of motion alone cannot account for all the observable phenomena of the evolution of this universe. Suppose we measure, over long tick sequences, the relative frequency with which the state of the system is less than $\frac{1}{2}$. Naturally we expect the result to be very close to $\frac{1}{2}$. However, this prediction cannot be deduced from the law of motion, due to the simple fact that there are trajectories which satisfy the law of motion and that have a relative frequency different from $\frac{1}{2}$: consider, for instance, the trajectory generated by the initial condition $\frac{1}{7}$, which determines a relative frequency of $\frac{2}{3}$. In order to obtain a given relative frequency, and particularly the frequency $\frac{1}{2}$, we need to define a measure on the trajectories which says that there are ``many more'' trajectories giving a relative frequency of $\frac{1}{2}$ than trajectories yielding other relative frequencies, and that therefore it is ``almost certain'' that the trajectory our universe is traveling along is one of those that yields a relative frequency of $\frac{1}{2}$.

Let us construct, therefore, a probability space for this universe. The space $S$ is the space of the trajectories of the system, the $\sigma$-algebra and the measure $\nu$ on $S$ derive from Borel's $\sigma$-algebra and from the Lebesgue measure of the interval $[0,1)$ through the one-to-one correspondence that exists between the two sets (see Appendix). Events are subsets of trajectories and can be described by statements such as: ``at the i-th tick, the state of the system is less than $\frac{1}{2}$,'' or: ``between the i-th and j-th ticks, the state of the system is greater than $\frac{1}{2}$''.

In order to establish a correspondence between probability and frequency, let us introduce the notion of sequence of equivalent events: we say that a sequence of events $E_1,...,E_n$ is a {\it sequence of equivalent events} if all the events have the same probability and are independent, i.e., $\nu (E_1\cap ...\cap E_n)=\nu (E_1)...\nu (E_n)$. For instance, if $E_i$ is the event ``at the i-th tick, the state of the system is less than $\frac{1}{2}$'', any sequence composed of events $E_i$ is a sequence of equivalent events with probability $\frac{1}{2}$. Given a sequence of equivalent events, we associate with every trajectory $\{a_i\}$ the number $\eta (\{a_i\})$ that indicates how many events of the sequence are verified by the trajectory (i.e., how many events it belongs to). By the law of large numbers, for any $\epsilon >0$,
\begin{equation} \label{33}
\nu \{\{a_i\}\in S |\epsilon\leq|\eta(\{a_i\})/n-P|\}\rightarrow 0 \,{\rm for}\, n\rightarrow \infty ,
\end{equation}
where $n$ is the number of events in the sequence and $P$ is their probability. Expression (\ref{33}) says that if $n$ is large, then there are very ``few'' trajectories which determine, for the relative frequency of events, a value which is significantly different from their probability. Thanks to espression (\ref{33}), one can therefore set the following correspondence between probability and frequency:
\begin{proposizione} \label{p:1} \rm The relative frequency with which the events of a long sequence of equivalent events occur corresponds to their probability \footnote{\label{note3} Actually, the passage from espression (\ref{33}) to Proposition \ref{p:1} is based on the implicit assumption that an event having very small probability does not occur. This assumption is accepted without discussing its validity or its conceptual implications.}.
\end{proposizione}
From this statement and from the definition of the events $E_i$ one easily obtains the value $\frac{1}{2}$ for the relative frequency described at the beginning of the section.

\subsubsection{Probability space and measurement theory for the asymptotic model.} \label{s322}

Let us now build, by analogy with the Bernoulli system, a probability space for the asymptotic model. The sample space is the set $B_C$ of classical N-bigbangs. The $\sigma$-algebra and the measure are transferred to $B_C$ from $V^N$ through the application $\omega _C:=\omega |_{B_C}:B_C\rightarrow V^N$. The following are defined on $V^N$: (i) Borel's $\sigma$-algebra ${\cal B}$ and the quantum measure $\mu_Q$, and (ii) the quotient $\sigma$-algebra $\tilde{\cal B}$ and the quotient measure $\tilde{\mu}_Q$ (see section \ref{s232} and the Appendix). The $\sigma$-algebra on $B_C$ must take into account statement 2b of the asymptotic model; therefore two disjoined events, i.e., two distinguishable events, which contain two N-bigbangs belonging to the same class of $\omega E$-equivalence will not be allowed. This requirement is certainly satisfied if the $\sigma$-algebra on $B_C$ is derived from the $\sigma$-algebra $\tilde{\cal B}$, because in this way the atoms of the derived $\sigma$-algebra are the classes of $\omega E$-equivalence themselves. The ``probability'' space that represent the universe will therefore be the tern
\begin{equation} \label{34}
(B_C,\tilde{\cal F},\tilde{\nu}_Q),
\end{equation}
where $\tilde{\cal F}:=\omega _C^{-1}(\tilde{\cal B})$, and $\tilde{\nu}_Q:=\omega _C^{-1}(\tilde{\mu }_Q)$ \footnote{ Actually, this method for transferring the measure $\mu_Q$ from $V^N$ onto $B_C$ or onto $V_I^N$ does not give correct results at the discontinuity points of the function $\omega_V$. See Appendix A for more details. I preferred to use this method anyway in order to simplify the exposition, but the reader should bear in mind that this method should be replaced everywhere with the method described in Appendix A.}. Note that if the $\sigma$-algebra on $B_C$ were derived from ${\cal B}$ rather than from $\tilde{\cal B}$, one would eliminate the fact that two classical N-bigbangs connected by a Galilean transformation are indistinguishable but not the fact that two N-bigbangs having the same asymptotic velocity are indistinguishable; the latter feature is {\it structural} to the asymptotic model and cannot be eliminated.

Actually, the tern (\ref{34}) does not represent a true probability space, since the measure $\tilde{\nu}_Q$ is not bounded, and therefore it cannot be normalized (one should bear in mind that there is no limit to the energy of the N-bigbangs of $B_C$). This fact will not prevent the construction of a consistent measurement theory: as we shall see, the probability of the result of a measurement will be defined as a ratio of measures.

In order to build a measurement theory for the asymptotic model (to which we will refer as {\it asymptotic theory of measurement}), consider two kinds of events: the {\it experiment} event and the {\it result} event. The experiment-event includes all those trajectories of $B_C$ which are macroscopically compatible with the preparation of a precise experiment occurring in a given space-time region. The result event is a subset of the experiment event, containing all the trajectories compatible with a certain result of the experiment. It is natural to define the probability $P(R|E)$ of a result $R$ of the experiment $E$ as
\begin{equation} \label{35}
P(R|E):=\frac{\tilde{\nu}_Q(R)}{\tilde{\nu}_Q(E)}.
\end{equation}
As in the previous example, it is not difficult to establish a correlation between probability and frequency: given an experiment $E$, we say that a sequence of results $R_1,...,R_n$ of the experiment is equivalent if $\tilde{\nu}_Q(R_1)=...= \tilde{\nu}_Q(R_n)$, and furthermore $\tilde{\nu}_Q(R_1\cap...\cap R_n)= \tilde{\nu}_Q(R_1)... \tilde{\nu}_Q(R_n)$. For instance, if the experiment consists of the emission of a large number of particles toward a screen, equivalent results are those in which different particles hit the same region of the screen. In this case also, the law of large numbers allows one to deduce that if $n$ is large, the probability that the relative frequency of the results $R_i$ differs significantly from their probability is almost null.

One should bear in mind that in the asymptotic model equation (\ref{35}) determines the probability of the results of {\it all} experiments of a statistical nature; for instance, it determines both the probability that tossing a coin will produce a given result and the probability that in the two-slit experiment a particle will hit the screen at a given point. We will see, in the latter case, how the phenomenon of the nonchaotic distribution of initial conditions allows to account for the quantum effect of interference fringes even within a classical dynamical context.


\subsubsection{Comparison between asymptotic and quantum measurement theory.}

The aim of this section is to compare asymptotic theory with the quantum theory of measurement, showing that if one makes a suitable assumption regarding the quantum measurement process they are compatible. This section is mathematically less rigorous out of necessity and for the sake of simplicity.

As a first simplification, we will consider as a probability space for the asymptotic model the tern
\begin{equation} \label{36}
(C,{\cal F},\nu_Q)
\end{equation}
instead of the tern (\ref{34}), where ${\cal F}:=\omega _C^{-1}({\cal B})$ and $\nu_Q:=\omega_C^{-1}(\mu_Q)$ . Hence the requirement that two N-bigbang connected by a Galilean transformation be indistinguishable has been disregarded. Therefore equation (\ref{35}) becomes:
\begin{equation} \label{37}
P(R|E)=\frac{\langle{\bf x}_N|E^+[\omega_C(R)]|{\bf x}_N\rangle}{\langle{\bf x}_N|E^+[\omega_C(E)]|{\bf x}_N\rangle}.
\end{equation}

As regards quantum theory, we will use the Everett-DeWitt Many-Worlds theory \cite{everett} for the comparison, as it is the most convenient for this purpose. In this theory, the state of the entire universe is represented by a wave function $\psi (x,t)$, and by obvious analogy with relation (\ref{16}) we assume that $\psi (x,0)=|{\bf x}_N\rangle$ . During evolution, every time a measurement (or an equivalent process) takes place, the state of the universe subdivides into branches which differ macroscopically from one another. Consider a measurement made at the time $t$ which involves $n$ possible distinct results; let $\psi$ be the state of the universe at the time of the measurement, and let $\psi_1,...,\psi_n$ be the states into which it decomposes. Obviously the equation $\sum_i\psi_i=\psi$  holds. The quantum probability $P_i$ of the i-th result is given by the classical formula
\begin{equation} \label{38}
P_i=\frac{|\langle\psi|\psi_i\rangle|^2}{\langle\psi|\psi\rangle\langle\psi_i|\psi_i\rangle}.
\end{equation}
One can achieve a tight link between the two measurement theories by making a conjecture for which we prepare a few definitions. Let the {\it space support} of a state $\psi$ be the support of the measure $\langle \psi |E^Q(\cdot)|\psi \rangle$ , and let the {\it asymptotic support} be the support of the measure $\langle \psi |E^+(\cdot)|\psi \rangle$ . We say that two states $\psi _1$ and $\psi _2$ are {\it spatially (asymptotically) disjoined} if their space (asymptotic) supports are disjoined. Finally, we say that the two states are {\it definitively disjoined} if $e^{-iHt}\psi _1$ and $e^{-iHt}\psi _2$ are spatially disjoined for every $t\geq 0$. Let us now make the following assumption:

\begin{proposizione} \label{p:2} \rm
The states into which the wave function of the universe subdivides due to a measurement are definitively disjoined.
\end{proposizione}
The fact that states that correspond to different results of a measurement are disjoined is widely accepted in the literature \cite{disgiunti}. This conclusion trivially derives from the fact that the two states must represent measurement instruments whose pointers occupy different positions (note that here, too, in order for two states of the universe to be spatially disjoined it is sufficient for them to describe a single particle in two different positions). The fact they are also definitively disjoined can be justified by a line of argument which is similar to the one used in section \ref{s31} to justify the fact that two macroscopically distinct trajectories have distinct asymptotic velocities. An assumption perfectly analogous to Proposition \ref{p:2} was made by Bohm \cite{bohm}.

Unfortunately, after presenting the physical reasons why Proposition \ref{p:2} must hold, we must acknowledge that from a mathematical standpoint, at least in the context developed in this paper, the proposition is impossible. In fact, it is well-known that a wave function that evolves according to Schr\''{o}dinger's equation and has a compact space support at a given instant, is spread to the entire space at any subsequent instant. Proposition \ref{p:2} can therefore be at most ``almost true''. We will replace it with the following weaker but more precise assumption:
\begin{proposizione} \label{p:3} \rm
The states into which the wave function of the universe subdivides due to a measurement are asymptotically disjoined.
\end{proposizione}
From Proposition \ref{p:3} and from the definition of asymptotic support one can derive that $\psi_i=E^+(\Delta_i)\psi$ , where $\Delta_i$ is the asymptotic support of $\psi_i$. In turn, the state $\psi$  derives from an alternate sequence of free evolutions and subdivisions
$$ \label{39}
\psi = e^{-iH(t-t_n)}E^+(\Delta^{(n)}) e^{-iH(t_n-t_{n-1})}...E^+(\Delta^{(1)}) e^{-iHt_1}|{\bf x}_N\rangle,
$$
where $\Delta^{(k)}$ is the asymptotic support that corresponds to the branching which took place at the k-th instant. Since the projectors $E^+(\Delta^{(k)})$ commute with the Hamiltonian, one obtains
\begin{equation} \label{40}
\psi =E^+(\Delta^{(n)}\cap ...\cap \Delta^{(1)}) e^{-iHt}|{\bf x}_N\rangle =E^+(\Delta_\psi ) e^{-iHt}|{\bf x}_N\rangle ,
\end{equation}
where $\Delta_\psi :=\Delta^{(n)}$ is the asymptotic support of $\psi$ . By replacing equation (\ref{40}) in equation (\ref{38}) one obtains
\begin{equation} \label{41}
P_i=\frac{\langle{\bf x}_N|E^+(\Delta_i)|{\bf x}_N\rangle}{\langle{\bf x}_N|E^+(\Delta_\psi)|{\bf x}_N\rangle}.
\end{equation}
This exactly corresponds to equation (\ref{37}), provided that one identifies the asymptotic supports of the branches into which the wave function subdivides with the sets of asymptotic velocities of the experiment events and result events of the asymptotic theory of measurement.


\subsection{Nonchaotic distribution of initial conditions.} \label{s33}

An important difference between the measure on the trajectories of the Bernoulli system and the measure on the N-bigbangs of the asymptotic model is that the former derives from a measure defined on the initial conditions, while the latter derives from a measure defined on the asymptotic velocities, i.e., on the final conditions. This feature admits, in the asymptotic model, the phenomenon of {\it nonchaotic distribution of initial conditions} (NCDIC), i.e., a distribution of the initial conditions that depends on the future evolution of the trajectories. The physical understanding of this extremely counterintuitive phenomenon is fundamental, in order to understand how typically quantum-like phenomena such as the two-slit experiment can be explained in a context where particle motion is described by classical trajectories. We will also see how this phenomenon is tightly linked to the EPR paradox. The phenomenon of NCDIC is remarkably similar to the phenomenon of preinteractive correlations, on which Price \cite{price} has written extensively.

One can illustrate this phenomenon in general in the following way. The law of motion allows to determine univocally the asymptotic velocity of a classical N-bigbang starting from its initial velocity, i.e., from the velocities of the particles at the origin of the N-bigbang. Let $V_I^N=R^{3N}$ be the space of initial velocities and let $\omega_V:V_I^N\rightarrow V^N$ be the application that describes this correspondence. We assume without demonstration that $\omega_V$ is measurable. On $V_I^N$ we define the measure $\pi _C$ (where $C$ stands for ``classical'') by setting
\begin{equation} \label{42}
\pi_C(\Delta ):=\frac{\mu_L(m\Delta )}{(2\pi )^{3N}},
\end{equation}
where $\mu_L$ is the Lebesgue measure. The measure $\pi _C$ describes a uniform (or {\it chaotic}) distribution of the initial momenta [the factor $(2\pi )^{-3N}$ is introduced for consistency with the quantum measure in the case of lack of potential, see equation (\ref{ar2})]. On $V_I^N$ one can also define the quantum measure $\pi_Q$ by deriving it through the application $\omega_V$ from the quantum measure $\mu_Q$ on the asymptotic velocities:
\begin{equation} \label{43}
\pi_Q:=\omega_V^{-1}(\mu_Q).
\end{equation}
The measure $\pi_Q$ is defined on the $\sigma$-algebra ${\cal B}_\omega :=\omega_V^{-1}({\cal B})$, which in general is different from ${\cal B}$; since $\omega_V$ is Borel-measurable, we have ${\cal B}_\omega \subseteq {\cal B}$. One should also consider note 4. The measure $\pi_Q$ describes the distribution that the initial velocities of the particles must have in order for the distribution of their asymptotic velocities to be $\mu_Q$. In general, one will have
\begin{equation} \label{44}
\pi_Q(\Delta )\neq \pi_C(\Delta ), \,\Delta \in {\cal B}_\omega.
\end{equation}
Formula (\ref{44}) expresses the fact that the asymptotic model has a {\it nonchaotic} distribution of the initial momenta of the particles. The structure of $\pi_Q$ will be extremely complex and intricate, because a minute difference in the initial velocities entails a large difference in the final asymptotic velocities.

The measure $\mu_C$, defined by
\begin{equation} \label{45}
\mu_C:=\omega_V(\pi_C),
\end{equation}
instead describes the distribution of asymptotic velocities that one would obtain if the distribution of initial momenta were uniform. The Appendix shows an equation which compares, albeit in an approximate way, the measures $\mu_C$ and $\mu_Q$.

We will now show how the possibility of an NCDIC radically modifies the physical interpretation of some major quantum experiments.


\subsubsection{The scattering process.}

Consider a flux of particles of the same type and with the same velocity oriented toward the positive direction of the $z$-axis and headed toward a target represented by a central potential $V(r)$. Let $\Theta$ be the scattering angle and let $\phi$  be the angle that the plane of the trajectory forms with the $x$-axis. The scattering cross-section $\sigma (\Theta ,\phi )$ is defined as
\begin{equation} \label{46}
\sigma(\Theta,\phi) \,{\rm d}\Omega=\frac{{\rm number\>of\> particles\> scattered\> for\> unit\> time\> in\> the\> solid\> angle\> d\Omega}}{{\rm \> incoming\> flux \> intensity}},
\end{equation}
where ${\rm d}\Omega$  is the solid angle $\sin\Theta \,{\rm d}\Theta \,{\rm d}\phi$, and the incoming flux intensity is the number of incoming particles per unit time divided by the total impact area, which we assume to be finite yet as large as one chooses.

The calculation of the classical cross-section is based on the statistical hypothesis that the flux of incoming particles is uniform in the incidence plane. Let us now develop a derivation of the cross-section which highlights this hypothesis, keeping it separate from the dynamical part of the calculation.

The trajectory of a particle is univocally determined by the impact parameter $s$ and by the angle $\phi$. For the time being, we are not making any hypotheses on the distribution of the incoming flux and we describe it generically by means of a density $\rho_I(s,\phi )$, so that if  $A$ is a region in the incident plane, the quantity $\int_A{\rho_I(s,\phi )s\,\,{\rm d}s\,{\rm d}\phi}$ is the number of particles per unit time that crosses the region $A$. Likewise, let $\rho_S(\Theta ,\phi )$ be the density of the scattered particles, so that $\int_\Omega{\rho_S(\Theta ,\phi )\sin\Theta \,{\rm d}\Theta \,{\rm d}\phi}$ is the number of particles scattered in the unit time within the solid angle $\Omega$. We will therefore have $\sigma (\Theta ,\phi )=\rho_S(\Theta ,\phi )/I$, where $I$ is the incoming intensity.

In order to make the calculation more realistic, let us suppose that the flux of incoming particles originates from a point-like source, located on the $z$-axis at the point $z_0$, which is at a great distance from the target. All the particles are emitted with the same energy and so that the direction of the velocities is distributed according to a given distribution $\rho_E(\theta,\phi )$, where $\theta$ is the angle formed with the $z$-axis. Obviously, the equation $s=\theta|z_0|$ holds, so that $\rho_I(s,\phi )= \rho_E(s/|z_0|,\phi )$. Therefore the distribution $\rho_I(s,\phi )$ derives from the minute fluctuations of the emission angle $\theta$, which are completely beyond the control of the experimenter.

The dynamics allows to express $\Theta$  as a function of $s$; for the sake of simplicity, let us assume that this correspondence is invertible, as certainly happens in the case of a repulsive potential. Clearly, 
$$ \label{47}
\rho_S(\Theta,\phi)\sin\Theta \,{\rm d}\Theta \,{\rm d}\phi=\rho_I(s,\phi)s\,\,{\rm d}s\,{\rm d}\phi=\rho_E(s/|z_0|,\phi)s\,\,{\rm d}s\,{\rm d}\phi
$$
holds. Since ${\rm d}s=|\partial s/\partial\Theta|{\rm d}\Theta$, we have:
\begin{equation} \label{48}
\rho_S(\Theta,\phi) =\rho_I(s,\phi)\frac{s}{\sin\Theta}\left|\frac{\partial s}{\partial\Theta}\right|.
\end{equation}

Equation (\ref{48}) is purely dynamical, i.e., it does not have any embedded statistical hypothesis. The classical cross-section is obtained by using the statistical hypothesis that the flux is uniform, i.e.,
\begin{equation} \label{49}
\rho_I(s,\phi )=\rho_E(s/|z_0|,\phi )=I.
\end{equation}
Hypothesis (\ref{49}) is normally considered so obvious that it is very difficult to find in the literature a line of reasoning which supports it or justifies it. It exactly corresponds to the hypothesis that the distribution of the velocities of the particles emitted by the source is chaotic and therefore does not depend on the subsequent interactions of the particles.

It is well-known that there are potentials for which the classical cross-section is not correct. It is therefore evident that either equation (\ref{48}) or hypothesis (\ref{49}) are not correct. The current interpretation states that equation (\ref{48}), i.e., the dynamics, is incorrect, and that it must be replaced with a quantum dynamics, while hypothesis (\ref{49}) remains substantially true. The interpretation that derives from the asymptotic model instead states the opposite, i.e., that equation (\ref{48}) is true and hypothesis (\ref{49}) is false. In the asymptotic model, it is not the distribution $\rho_S$ that derives from the distribution $\rho_E$ through equation (\ref{48}), but vice versa it is the distribution $\rho_E$ that, through the same equation, derives from $\rho_S$, which in turn derives from the measure on the asymptotic velocities of the particles of the universe [note the analogy of this procedure with the one used in the previous section to derive the measure $\pi_Q$ from the measure $\mu_Q$; in this example, the application $\Theta(s)$ plays the role of the application $\omega_V$]. This reversal of perspective renders the distribution of the velocities with which particles are released by the source nonchaotic and allows it to depend on the form of the potential.


\subsubsection{The two-slit experiment.} \label{s332}

Particle diffraction in the two-slit experiment is probably, along with the tunnel effect, the phenomenon that best convinced physicists that they needed to abandon classical trajectories to describe particle motion. See for instance Heisenberg \cite{heisenberg}. However, we see now that NCDIC allows an interpretation of this experiment which does not entail renouncing classical trajectories.

Consider the experimental apparatus of Figure 1, which is the apparatus by which the two-slit experiment with electrons has been actually performed.

\begin{center}
\unitlength=1mm
\begin{picture}(120,35)
\put(54,5){\line(1,0){10}}
\put(54,30){\line(1,0){10}}
\put(58,17){\circle*{1}}
\put(112,5){\line(0,1){23}}
\put(0,16){\line(1,0){7}}
\put(7,16){\line(0,1){3}}
\put(0,19){\line(1,0){7}}
\put(0,16){\line(0,1){3}}
\bezier{600}(7,18)(58,22)(112,17)
\bezier{600}(7,17)(58,13)(112,17)

\put(47,2){\makebox(6,6){$E_2$}}
\put(47,27){\makebox(6,6){$E_1$}}
\put(50,14){\makebox(6,6){$F$}}
\put(105,1){\makebox(6,6){$H$}}
\put(112,15){\makebox(6,6){$P$}}

\put(0,10){\makebox(6,6){$S$}}

\end{picture}

Fig. 1
\end{center}


Here $S$ is an electron source, $F$ is a tiny conducting wire which crosses the plane of the figure at right angles and is set to a positive electric potential with respect to the two electrodes $E_1$ and $E_2$; $H$ is a screen constituted by a photographic plate. Due to the electrostatic field generated by the wire, the electrons emitted by the source are deflected and produce interference fringes on the screen. If the electrostatic field is turned off, the interference fringes disappear.

This experiment is completely equivalent to the scattering experiment: there is a distribution $\rho_E$ of the directions of the velocities with which electrons are emitted by the source, and there is a distribution $\rho_S$ of the locations where they hit the plate. If one maintains the assumption that between the source and the screen the particles travel along classical trajectories, the two distributions $\rho_E$ and $\rho_S$ are mutually linked by a dynamical equation which is analogous to equation (\ref{48}). The paradox arises from the implicit assumption that the distribution that ``controls'' the experiment is $\rho_E$, that it is chaotic, and that therefore it cannot depend on the fact that an electrostatic field is acting or not. All paradoxes disappear, however, if one admits that the ``controlling'' distribution is $\rho_S$, which is derived from the quantum measure of the asymptotic boundary conditions of the universe, and that the distribution $\rho_E$ derives from it.

A typical question regarding the two-slit experiment is the following: through which of the two slits (in this case, on which side of the wire) does the particle pass? Let us see how the asymptotic model answers this question.

The figure shows the two classical trajectories which make the particles hit the plate at a same point $P$. Those two trajectories, intended as overall trajectories of the universe, do not differ macroscopically from one another, since the electrons darken the same silver grain on the plate. One can therefore reasonably assume that they have the same asymptotic velocity. Therefore the two trajectories correspond to two different classical reference frames of the same N-bigbang on $M$, which is the actual real physical object. In this sense one can state that the particle travels through both slits.


\subsubsection{The EPR paradox.}

The demonstration of Bell's inequality also implicitly uses the hypothesis that the distribution of initial conditions is chaotic. Without this hypothesis, Bell's inequality, and therefore the EPR paradox, can no longer be demonstrated. Let us now show where, in Bell's classical demonstration \cite{bell}, this hypothesis is used.

A source emits pairs of particles in opposite directions towards two measurement instruments, each able to measure two distinct observables. In total, therefore, one can make four different measurements on each pair of particles. On the basis of the well-known criteria of local realism, one can deduces that every pair can be assigned, when it is emitted by the source, a hidden variable $\lambda$ which univocally determines the result of each one of the four measurements. The demonstration considers the execution of four distinct experiments: in each one, one of the four measurements is made on N pairs. Let $\rho_i(\lambda), \, i=1,...,4$ be the distributions of the variable $\lambda$ in the four experiments. Bell derives his inequality by assuming that the distributions $\rho_i(\lambda)$ are the same in the four experiments. This is exactly the chaotic hypothesis. If the chaotic hypothesis is not used, one can admit that the distributions $\rho_i(\lambda)$ depend on the measurement made, in the same way in which in scattering the distribution of the incoming particles can depend on the form of the potential. It is straightforward to prove that without the condition that the distributions $\rho_i(\lambda)$ are equal to one another the inequality can no longer be demonstrated.

Despite this conclusion, the asymptotic model cannot describe the EPR experiment in terms of classical trajectories, because it cannot deal with particles with spin. This result could possibly be attained by extending the model.


\subsubsection{Atomic levels and the tunnel effect.} \label{s334}

It is easy to realize that from a formal standpoint the bare Schr\"{o}dinger equation cannot explain atomic levels. In fact it does not require an orbital electron to be in an eigenstate of the Hamiltonian, nor does it allow the electron to decay from one energy level to another. In order to achieve a more accurate representation of these phenomena one needs to add to the Hamiltonian a term which represents interaction with the electromagnetic field. The asymptotic model, by construction, inevitably shares this limitation with Schr\"{o}dinger's equation. An asymptotic model more suitable to represent those phenomena could possibly be obtained by extending this model to quantum field theory.

Let us now consider the tunnel effect. Consider for instance an $\alpha$-decay, where an atom of radium emits an $\alpha$ particle and becomes an atom of radon. The ideal representation of this phenomenon describes the $\alpha$ particles as being confined in a potential well, as in Figure,

\begin{center}
\unitlength=1mm
\begin{picture}(50,18)
\put(0,0){\line(1,0){50}}
\put(20,0){\line(0,1){15}}
\put(30,0){\line(0,1){15}}
\bezier{600}(0,1)(18,3)(20,15)
\bezier{600}(30,15)(32,3)(50,1)
\put(12,13){\makebox(5,5){$V_p$}}
\end{picture}

Fig. 2
\end{center}
which is determined by the short-range nuclear attractive forces and by the weaker long-range Coulomb repulsive force. The kinetic energy of the emitted $\alpha$  particles is lower than the peak potential $V_p$; this is classically considered impossible.

Indeed, even according to the asymptotic model, in a universe consisting of a single $\alpha$ particle subjected to the potential of Figure 2 the tunnel effect would not be observable. In fact, by using a line of argument similar to the one used in section \ref{s221}, one can demonstrate that by setting as asymptotic boundary condition a velocity less than $\sqrt{2V_p/m_\alpha}$  one obtains for the $\alpha$  particle the limit trajectory which at the time $t=+\infty$  reaches the peak of the potential.

However, in a real experiment, the situation is much more complex: the $\alpha$  particle interacts with the other particles in the atomic nucleus; furthermore, around the nucleus there are orbital electrons and finally the atom interacts with other atoms within the fissionable material. It is therefore possible for a suitable concurrence of this complex set of interactions to produce a classical trajectory which is compatible with the observed behavior. For instance, an orbital electron suitably hit by other orbital electrons or by a nearby atom could come close enough to the nucleus to lower the peak potential and allow the escape of the $\alpha$  particle having the required energy.

This ``conspiracy'' of interactions appears absurd and unacceptable only if one does not take into account that here, too, there is the same reversal of perspective mentioned earlier during the analysis of the scattering experiment and of the two-slit experiment: it is not the distribution of the initial conditions that rules the experiment, but rather the distribution of the asymptotic conditions; this distribution entails high probabilities for asymptotic conditions that correspond to the tunnel effect and the trajectories conform to those conditions, no matter how particular and ``conspiratorial'' the interactions taking place during the evolution might appear. One has to take into account, however, that this representation of the tunnel effect by the asymptotic model is unavoidably affected by its inadequacy in representing atomic phenomena, as explained at the beginning of the section.

The compatibility of the tunnel effect with classical trajectories was also stated in \cite{tunnel}.

\section{Conclusion.}
In order to understand whether the approach proposed in this paper is valid and fruitful, one must succeed in extending it to curved space-time and to quantum field theory. In my opinion the connection it highlights between a class of space-time transformations and a very general quantization mechanism such as Feynman path integrals makes it worthwhile to proceed with research in this direction.

\appendix
\section*{Appendix}

{\bf Transfer of measures}. Let $A$ and $B$ be two sets and let $f:A\rightarrow B$ be an application. We describe two ways of transferring $\sigma$-algebras and measures between $A$ and $B$:
\begin{itemize}
\item Let a $\sigma$-algebra ${\cal F}_B$ and a measure $\mu_B$ be defined on $B$. They can be transferred on $A$ by defining ${\cal F}_A:=f^{-1}({\cal F}_B):=\{f^{-1}(\Delta_B)|\Delta_B\in {\cal F}_B\}$ and $\mu_A(\Delta_A):=[f^{-1}(\mu_B)](\Delta_A):=\mu_B[f(\Delta_A)], \, \Delta_A\in {\cal F}_A$.
\item  Let two $\sigma$-algebras ${\cal F}_A$ and ${\cal F}_B$ be defined on $A$ and $B$, respectively, let the measure $\mu_A$ be defined on $A$ and let the application $f$ be measurable, i.e., $\Delta_B\in {\cal F}_B$ implies $f^{-1}(\Delta_B)\in {\cal F}_A$. The measure $\mu_A$ can be transferred on $B$ by defining $\mu _B(\Delta _B):=[f(\mu _A)](\Delta _B):=\mu _A(f^{-1}[\Delta _B)]$.
\end{itemize}
Notice that if $A$ and $B$ have the $\sigma$-algebras ${\cal F}_A$ and ${\cal F}_B$, and $f$ is measurable, then $f^{-1}({\cal F}_B)\subseteq {\cal F}_A$.
\vspace{4mm} \\
{\bf Transfer of the measure $\mu_Q$}. In section \ref{s33}, the measure $\mu _Q$ was transferred from $V^N$ onto $V_I^N$ by means of the application $\omega _V$, by defining
\begin{equation} \label{50}
\pi_Q := \omega^{-1}_V(\mu_Q).
\end{equation}
This definition is not correct for sets containing discontinuity points of the function $\omega_V$. The correct definition is the following: consider the application $X_t:V_I^N\rightarrow R^{3N}$ which associates with a velocity $v\in V_I^N$ the position at time t of the particles of a classical N-bigbang with origin at $(0,{\bf 0})$ and initial velocity $v$. We furthermore indicate by $\eta_{Qt}$ the measure
\begin{equation} \label{53}
\eta_{Qt}(\Delta_X)=\langle {\bf 0}_N|e^{iHt}E^Q[\Delta_X]e^{-iHt}|{\bf 0}_N\rangle,\, \Delta_X\subseteq R^{3N},
\end{equation}
hence
\begin{equation} \label{54}
\mu_Q(\Delta)=\lim_{t\rightarrow+\infty}\eta_{Qt}[\Delta^c (t)], \,\Delta \subseteq V^N.
\end{equation}
The measure $\pi_Q$ should be defined as
\begin{equation} \label{55}
\pi_Q(\Delta_I):= \lim_{t\rightarrow+\infty}\eta_{Qt}[X_t(\Delta_I)], \,\Delta_I \subseteq V_I^N.
\end{equation}
One can easily appreciate the difference between the definitions (\ref{50}) and (\ref{55}) by applying them to the example in section \ref{s221}. Choosing $b>0$, we set $\Delta_b:=[-b,b]\subseteq V_I$; the interval $\Delta_b$ embeds the point $v=0$ which is a discontinuity point for the function $\omega_V$ defined by (\ref{11}). Since
$$ \label{57}
\omega_V(\Delta_b)=\left[-\sqrt{2V_0/m+b^2},-\sqrt{2V_0/m}\right]\cup\{0\}\cup \left[\sqrt{2V_0/m},\sqrt{2V_0/m+b^2}\right].
$$
by using definition (\ref{50}) one has
$$ \label{58}
\pi_Q(\Delta_b)=\mu_Q\left(\left[-\sqrt{2V_0/m+b^2},-\sqrt{2V_0/m}\right]\cup\{0\}\cup \left[\sqrt{2V_0/m},\sqrt{2V_0/m+b^2}\right]\right).
$$
On the other hand, since for $t\gg 0$ one has
$$ \label{59}
X_t(\Delta_b)\cong\left[-t\sqrt{2V_0/m+b^2},t\sqrt{2V_0/m+b^2}\right],
$$
by using the second definition one obtains
$$ \label{60}
\pi_Q(\Delta_b)=\mu_Q\left(\left[-\sqrt{2V_0/m+b^2},\sqrt{2V_0/m+b^2}\right]\right).
$$

The application $\alpha :B_C\rightarrow V_I^N$ that associates with every classical N-bigbang its own initial velocity allows to define on $B_C$ the measure $\nu_Q$:
\begin{equation} \label{61}
\nu_Q := \alpha^{-1}(\pi_Q). 
\end{equation}
Equations (\ref{61}) describe the correct definition of $\nu_Q$, to be used instead of the one of section \ref{s322}. The quotient measure $\tilde{\nu}_Q$ is defined in the same way.
\vspace{4mm} \\
{\bf Quotient Measure}. Let $A$ be a set equipped with a $\sigma$-algebra ${\cal F}$ and with a measure $\mu$, on which there acts a group of transformations $G$, and let the measure $\mu$  be invariant under such transformations, i.e., $\mu(\Delta )=\mu [g(\Delta )], \,g\in G$. The relation of $G$-equivalence is defined on $A$, so that $a_1,a_2\in A$ are $G$-equivalent if there exists a $g\in G$ such that $a_2=g(a_1)$. Let $\tilde{A}$ be the quotient space of this relation. The $\sigma$-algebra $\tilde{{\cal F}}$ is defined for $\tilde{A}$ by setting $\tilde{{\cal F}}:={\cal P}(\tilde{A})\cap {\cal F}$, where ${\cal P}(\tilde{A})$ is the set of parts of $\tilde{A}$. The $\sigma$-algebra $\tilde{{\cal F}}$ is formed by the ${\cal F}$-measurable sets that are invariant under the transformations of the group G, and can also be thought as a $\sigma$-subalgebra of ${\cal F}$. The aim here is to define on $\tilde{A}$ a measure $\tilde{\mu}$ which corresponds to the measure $\mu$ on $A$.

Let $B\subseteq A$ be an ${\cal F}$-measurable set which represents the set $\tilde{A}$, i.e., which contains one and only one element for every $G$-equivalence class, and let $h:\tilde{A}\rightarrow B$ be the corresponding projection. The application $k:B\times G\rightarrow A$ is defined by setting $k(a,g):=g(a)$. By Haar's theorem \cite{haar}, on $G$ there exists a single measure $\mu_H$ (defined modulo a proportionality factor) which is left-invariant, i.e., such that $\mu_H(g\Delta_G)=\mu_H(\Delta_G)$ for every $\Delta_G\subseteq G$. Let us define measure $\tilde{\nu}$ on $B$ by setting
\begin{equation} \label{63}
\tilde{\nu}(\Delta_B)=\frac{\mu[k(\Delta_B\times\Delta_G)]}{\mu_H(\Delta_G)},
\end{equation}
where $\Delta_G\subseteq G$ is a set such that $0<\mu_H(\Delta_G)<+\infty$. One can prove that $\tilde{\nu}$ depends neither on the choice of the set $B$ nor on the choice of the set $\Delta_G$.

Finally, the {\it quotient} measure $\tilde{\mu}$ is defined by setting
\begin{equation} \label{64}
\tilde{\mu}:=h^{-1}(\tilde{\nu}).
\end{equation}
{\it Example}. Let $A=R^2$, let $\mu$  be Lebesgue's measure and let $G$ be the group of translations along $x:\, g(x,y)=(x+x_g,y)$. The elements of the quotient set $\tilde{A}$ are the straight lines that are parallel to the $x$-axis, while the elements of $\tilde{\cal F}$ are the measurable ``strips'' that are parallel to the $x$-axis. The set $B$ can be, for instance, the straight line $x=0$. The group $G$ is isomorphic to $R$, and its Haar measure is the Lebesgue measure on $R$. If $\Delta_B$ is a subset of $B$, for instance an interval, and $\Delta_G$ is another interval, then $k(\Delta_B\times\Delta_G)$ is a rectangle with base $\Delta_G$ and height $\Delta_B$, $\mu [k(\Delta_B\times\Delta_G)]$ its area, and $\tilde{\nu}(\Delta_B)$ is the length of the interval $\Delta_B$.
\vspace{4mm}\\
{\bf Comparison between classical and quantum measure}. The classical measure $\mu _C$ and the quantum measure $\mu_Q$ on $V^N$ [see equations (\ref{16}) and (\ref{45})] can be derived by taking to the limit $t\rightarrow +\infty$  the following two measures:
\begin{itemize}
\item Classical measure: $\eta_{Ct}(\Delta _X):=\pi_C[X_t^{-1}(\Delta_X)], \, \Delta_X\subseteq R^{3N}$, where $X_t:V_I^N\rightarrow R^{3N}$ is defined above in the appendix, and $\pi_C$ is the uniform measure on momenta defined by (\ref{42}).
\item Quantum measure: $\eta_{Qt}(\Delta _X):=\langle {\bf 0}_N|e^{iHt}E^Q[\Delta _X]e^{-iHt}|{\bf 0}_N\rangle, \, \Delta _X\subseteq R^{3N}$.
\end{itemize}
One can easily see that for both measures
\begin{equation} \label{65}
\mu_{C(Q)}(\Delta)=\lim_{t\rightarrow+\infty}\eta_{C(Q)t}[\Delta^c(t)], \, \Delta\subseteq V^N
\end{equation}
holds.

Let us indicate with $\rho_C(t,x)$ and $\rho_Q(t,x)$ the densities that correspond to the measures $\eta _{Ct}$ and $\eta_{Qt}$. In \cite{semiclassico} it is demonstrated that
\begin{equation} \label{66}
\rho_C(t,x)=\frac{1}{(2\pi)^{3N}}\sum\limits_{i}{\left|\det \left(\frac{\partial W_i}{\partial x_1\partial x_2}\right)\right|}(0,0,t,x),
\end{equation}
where $W_i(t_1,x_1,t_2,x_2)$ is the action of the $i$-th classical trajectory that joins the points $(t_1,x_1)$ and $(t_2,x_2)$. The same paper demonstrates that in a semi-classical approximation, $\rho_Q$ is given by
\begin{equation} \label{67}
\rho_Q(t,x)=\frac{1}{(2\pi)^{3N}}\left| \sum\limits_{i}{\left|\det \left(\frac{\partial W_i}{\partial x_1\partial x_2}\right)\right|^{1/2}\exp\left\{i\left(W_i-\frac{M_i\pi}{2}\right)\right\}}\right|^2(0,0,t,x),
\end{equation}
where $M_i$ is a phase factor. From (\ref{66}) and from (\ref{67}) one obtains
\begin{equation} \label{68}
\rho_C(t,x)=\rho_Q(t,x)+I(t,x),
\end{equation}
where $I(t,x)$ is the following interference term:
\begin{eqnarray} \label{ar4}
I(t,x)& = &\sum_{i\neq j} \left|\frac{\partial W_i}{\partial x_1\partial x_2}\right|^{1/2}\left|\frac{\partial W_j}{\partial x_1\partial x_2}\right|^{1/2}\times \nonumber \\
&& \times \exp\left[i\left((W_i-W_j)-\frac{(M_i-M_j)\pi}{2}\right)\right](0,0,t,x).
\end{eqnarray}


\end{document}